\documentclass[a4paper,11pt]{article}
\usepackage{subfig}
\usepackage{jheppub}
\usepackage{multirow}
\pdfoutput=1

\let\jnfont=\rm
\def\NPB#1,{{\jnfont Nucl.\ Phys.\ B }{\bf #1},}
\def\PLB#1,{{\jnfont Phys.\ Lett.\ B }{\bf #1},}
\def\EPJC#1,{{\jnfont Eur.\ Phys.\ Jour.\ C }{\bf #1},}
\def\PRD#1,{{\jnfont Phys.\ Rev.\ D }{\bf #1},}
\def\PRL#1,{{\jnfont Phys.\ Rev.\ Lett.\ }{\bf #1},}
\def\MPLA#1,{{\jnfont Mod.\ Phys.\ Lett.\ A }{\bf #1},}
\def\JPG#1,{{\jnfont J.\ Phys.\ G}{\bf #1},}
\def\CTP#1,{{\jnfont Commun.\ Theor.\ Phys.\ }{\bf #1},}
\def\ZPC#1,{{\jnfont Z.\ Phys.\ C }{\bf #1},}
\def\JHEP#1,{{\jnfont JHEP \ }{\bf #1},}

\title{Natural NMSSM after LHC Run I and the Higgsino dominated dark matter scenario}

\author[a,b]{Junjie Cao,}
\author[a]{Yangle He,}
\author[a]{Liangliang Shang,}
\author[c,d]{Wei Su,}
\author[c,d]{Yang Zhang}

\affiliation[a]{Department of Physics,
                Henan Normal University, Jianshe East Road, Xinxiang 453007, China}
\affiliation[b]{Department of Applied Physics, Xi'an Jiaotong University, Xianning West Road,
          Xi'an, Shanxi 710049, China }
\affiliation[c]{CAS Key Laboratory of Theoretical Physics, Institute of Theoretical Physics, Chinese Academy of Sciences, Zhong Guan Cun East Street, Beijing 100190, China}
\affiliation[d]{School of Physical Sciences, University of Chinese Academy, Yuquan Road, Beijing 100049, China}

\emailAdd{junjiec@itp.ac.cn}
\emailAdd{heyangle90@gmail.com}
\emailAdd{shlwell1988@gmail.com}
\emailAdd{weisv@itp.ac.cn}
\emailAdd{zhangyang@itp.ac.cn}

\abstract{We investigate the impact of the direct searches for SUSY at LHC Run I on the naturalness of the Next-to-Minimal Supersymmetric
Standard Model (NMSSM). For this end,  we first scan the vast parameter space of the NMSSM to get the region where the fine tuning measures
$\Delta_Z$ and $\Delta_h$ at the electroweak scale are less than about 50, then we implement by simulations the
constraints of the direct searches on the parameter points in the region. Our results indicate that although the direct search experiments are effective in excluding the points,
the parameter intervals for the region and also the minimum reaches of $\Delta_Z$ and $\Delta_h$ are scarcely changed by the constraints, which implies that
the fine tuning of the NMSSM does not get worse after LHC Run I. Moreover, based on the results we propose a natural NMSSM scenario where
the lightest neutralino $\tilde{\chi}_1^0$ as the dark matter (DM) candidate is Higgsino-dominated. In this scenario, $\Delta_Z$ and $\Delta_h$ may be as low as 2 without conflicting with any experimental constraints, and intriguingly $\tilde{\chi}_1^0$ can easily reach the measured DM relic density due to its significant Singlino component. We exhibit the features of the scenario which distinguish
it from the other natural SUSY scenario, including the properties of its neutralino-chargino sector and scalar top quark sector. We emphasize that the scenario can be tested either through searching for $3 l + E_T^{miss}$ signal at 14 TeV LHC or through future DM direct detection experiments. }
\begin{document}
\maketitle
\newpage

\section{Introduction}
\label{sec:introduction}

In the supersymmetric models such as the Minimal Supersymmetric Standard Model (MSSM) \cite{MSSM-1,MSSM-2} and the
Next-to-Minimal Supersymmetric Standard Model (NMSSM) \cite{NMSSM-1,NMSSM-2}, the $Z$ boson mass is given by
\cite{Baer:2012uy}
\begin{eqnarray}
m^2_{Z}=\frac{2 (m^2_{H_d}+\Sigma_{d})- 2 (m^2_{H_u}+
\Sigma_{u})\tan^{2}\beta}{\tan^{2}\beta-1}- 2 \mu^{2},
\label{minimization}
\end{eqnarray}
where $m^2_{H_d}$ and $m^2_{H_u}$ represent the weak scale soft SUSY
breaking masses of the Higgs fields $H_d$ and $H_u$ respectively,
$\Sigma_{d}$ and $\Sigma_{u}$ are their radiative corrections,
$\mu$ is the Higgsino mass and $\tan\beta\equiv v_u/v_d$. As was shown in \cite{Baer:2012cf},
the corrections $\Sigma_{d}$ and $\Sigma_{u}$ can be obtained from the effective Higgs potential
at loop level, and in case of a large $\tan \beta$, their largest
contributions arise from the Yukawa interactions
of third generation squarks, which are
\begin{eqnarray}
\Sigma_u &\simeq & \sum_{i=1}^2 \frac{3Y_t^2}{16\pi^2}\times m^{2}_{\tilde{t}_i}
\left( \log\frac{m^{2}_{\tilde{t}_i}}{Q^2}-1\right), \nonumber \\
\Sigma_d &\simeq & \sum_{i=1}^2 \frac{3Y_b^2}{16\pi^2}\times m^{2}_{\tilde{b}_i}
\left( \log\frac{m^{2}_{\tilde{b}_i}}{Q^2}-1\right).  \nonumber
\label{rad-corr}
\end{eqnarray}
In above formulae, $Q$ denotes the renormalization scale in getting the effective potential,
and its optimized value is usually taken as $Q = \sqrt{m_{\tilde{t}_1} m_{\tilde{t}_2}}$ with
$\tilde{t}_1$ and $\tilde{t}_2$ being the light and heavy top squarks (stop) respectively. Obviously,
if the observed value of $m_Z$ is obtained without
resorting to large cancelations, each term on the
right hand side of Eq.~(\ref{minimization}) should be comparable in
magnitude with $m_Z^2$, and this in return can put non-trivial constraints on the magnitudes of
$\mu$ and $m_{\tilde{t}_{1,2}}$. Numerically speaking, we find that requiring the individual term
to be less than $10  m_Z^2$ leads to $\mu$ and $m_{\tilde{t}_{1,2}}$ upper bounded by
about 200 GeV and 1.5 TeV respectively. In history, the scenario satisfying the bounds is dubbed as
Natural SUSY (NS) \cite{Baer:2012uy}.

In the MSSM, the NS scenario is theoretically unsatisfactory due to at least three considerations. First,
since the Higgsino mass $\mu$ is the only dimensionful parameter in the superpotential of the MSSM, its typical size
should be of the order of the SUSY breaking scale. Given that the LHC searches for supersymmetric particles have pushed the masses
of gluinos and first generation squarks up to above $1 {\rm TeV}$ \cite{ATLAS-Multi-jets, CMS-Multi-jets},
$\mu \lesssim 200 {\rm GeV}$ seems rather unnatural. Second, the relic density of the dark matter (DM) predicted in the NS
scenario is hardly to coincide with its measured value.  Explicitly speaking,
it has been shown that if the DM $\tilde{\chi}_1^0$ is Higgsino-dominated \footnote{Throughout this work, we denote
the mass eigenstates of the neutralinos by $\tilde{\chi}_i^0$ with $i$ ranging from 1 to 4 (5) for MSSM (NMSSM), and assume
an ascending mass order for the $\tilde{\chi}_i^0$ by convention.  With such an assumption,
the lightest neutralino $\tilde{\chi}_1^0$ as the lightest supersymmetric particle (LSP) is regarded as the DM candidate.},
its density is usually about one order smaller than its measured value \cite{Baer:2012uy}, alternatively if it is Bino-dominated, the correct density can be achieved
only in very limited parameter regions of the MSSM \cite{Cao:2015efs}. These features make the NS scenario disfavored by DM physics. Third, the NS scenario
is further exacerbated by the uncomfortably large mass of the recently discovered Higgs particle \cite{ATLAS-2012,CMS-2012}:
its value $m_h \simeq 125 {\rm GeV}$ lies well beyond its tree-level upper bound $m_h \leq m_Z$, and consequently stops heavier than about $1 {\rm TeV}$
must be present to provide a large radiative correction to the mass \cite{MSSM-Early-1,MSSM-Early-2,MSSM-Early-3,MSSM-Early-4,MSSM-Early-5,MSSM-Early-6}.
This requirement seems in tension with the naturalness argument
of Eq.~(\ref{minimization}).  In fact, all these problems point to the direction that the NS scenario should be embedded in a more complex framework.
Remarkably, we note that the NMSSM is an ideal model to alleviate these problems.

The NMSSM extends the MSSM by one gauge singlet superfield $\hat{S}$,
and it is the simplest SUSY extension of the Standard Model (SM) with a scale invariant superpotential (i.e. its superpotential does not
contain any dimensionful parameters) \cite{NMSSM-1,NMSSM-2}. In this model, the Higgsino mass $\mu$ is dynamically generated by the vacuum expectation value of
$\hat{S}$, and given that all singlet-dominated scalars are lighter than about $v \simeq 174 {\rm GeV}$, its magnitude can be naturally less than $200 {\rm GeV}$.
These additional singlet-dominated scalars, on the other hand, can act as the mediator or final states of the DM annihilation \cite{Cao:2015loa}, and consequently
the NS scenario in the NMSSM with a Singlino-dominated DM can not only predict the correct relic density, but also explain the galactic center
$\gamma$-ray excess \cite{Cao:2015loa,Cao:2014efa,Guo:2014gra,Bi:2015qva}. Moreover, in the NMSSM the interaction  $\lambda \hat{S} \hat{H}_u \cdot \hat{H}_d $
can lead to a positive contribution to the squared mass of the
SM-like Higgs boson, and if the boson corresponds to the next-to-lightest CP-even Higgs state, its mass can be further lifted up by the
singlet-doublet Higgs mixing. These enhancements make the large radiative correction of the stops unnecessary in predicting $m_h \simeq 125 {\rm GeV}$,
and thus stops can be relatively light \cite{NMSSM-Early-1,NMSSM-Early-2,NMSSM-Early-3,NMSSM-Early-4,NMSSM-Early-5,NMSSM-Early-6,NMSSM-Early-7,NMSSM-Early-8}.

So far studies on the NS scenario in the NMSSM are concentrated on the assumption that $\tilde{\chi}_1^0$ is Singlino-dominated \cite{Cao:2015loa,Cao:2014efa,Guo:2014gra,Bi:2015qva,
Singlino-NMSSM-1,Singlino-NMSSM-2,Singlino-NMSSM-3,Singlino-NMSSM-4,Singlino-NMSSM-5,Singlino-NMSSM-6,Singlino-NMSSM-7,Singlino-NMSSM-8,Singlino-NMSSM-9,
Singlino-NMSSM-10,Singlino-NMSSM-11,Singlino-NMSSM-12,Singlino-NMSSM-13,Singlino-NMSSM-14,Singlino-NMSSM-15,Singlino-NMSSM-16,Singlino-NMSSM-17}.
In this case, the branching ratio of the golden channel $\tilde{t}_1 \to t \tilde{\chi}_1^0$ in the LHC search for a moderately light stop
is highly suppressed.  Instead, $\tilde{t}_1$ mainly decays into the Higgsino-dominated $\tilde{\chi}_1^+$ and $\tilde{\chi}_{2,3}^0$ in following way \cite{Singlino-NMSSM-13}
\begin{eqnarray}
\tilde{t}_1 \to b \tilde{\chi}_1^{+} \to  b W^{+ (\ast)} \tilde{\chi}_1^0, \quad \tilde{t}_1 \to t \tilde{\chi}_{2,3}^0 \to  t X^{0 (\ast)} \tilde{\chi}_1^0, \label{Singlino-LSP}
\end{eqnarray}
where $X^0$ denotes either $Z$ boson or a neutral Higgs boson. These lengthened decay chains can generate softer final particles
in comparison with the golden channel, and consequently weaken the LHC bounds in the stop search.
This feature is also applied to other sparticle searches, and it has been viewed as an advantage of the NMSSM in circumventing the tight
constraints from the LHC searches for SUSY. In this work, we consider another realization of the NS scenario where $\tilde{\chi}_1^0$ is
Higgsino-dominated. In our scenario, the Higgsinos and the Singlino are degenerated in mass at $50\%$ level, and
consequently they mix rather strongly to form mass eigenstates $\tilde{\chi}_{1,2,3}^0$ with $\tilde{\chi}_{1,2}^0$
and $\tilde{\chi}_3^0$ being Higgsino-dominated and Singlino-dominated respectively.  Since the role of the Singlino component in $\tilde{\chi}_1^0$
is to decrease the DM annihilation rate,  $\tilde{\chi}_1^0$ may achieve the relic density measured by Planck and WMAP experiments \cite{Planck,WMAP}
without contradicting the DM direct search experiments such as LUX \cite{LUX,LUX-1}. The phenomenology of our scenario is somewhat similar to that
of the popular NS scenario in the MSSM, which was proposed in \cite{Baer:2012uy}, but our scenario has following advantages
\begin{itemize}
\item In the parameter regions allowed by current LHC searches for SUSY, it may have a lower fine tuning in getting the $Z$ boson mass. Meanwhile, it has broad parameter
regions to predict the right relic density (see our discussions in Sect. III).
\item The mass gaps of $\tilde{\chi}_1^\pm$ and $\tilde{\chi}_2^0$  from $\tilde{\chi}_1^0$ are sizable, e.g.
$\Delta_{\pm} \equiv m_{\tilde{\chi}_1^\pm} - m_{\tilde{\chi}_1^0} \gtrsim 30 {\rm GeV}$ and $\Delta_{0} \equiv m_{\tilde{\chi}_2^0} - m_{\tilde{\chi}_1^0}
\gtrsim 50 {\rm GeV}$, and consequently the leptons from the decay chains $\tilde{\chi}_1^\pm \to W^{(\ast)} \tilde{\chi}_1^0 \to l \nu \tilde{\chi}_1^0$
and $\tilde{\chi}_2^0 \to Z^{(\ast)} \tilde{\chi}_1^0 \to l l \tilde{\chi}_1^0$ are usually energetic. As a result, our scenario  can be tested at future LHC experiments
by the process $p p \to \tilde{\chi}_1^\pm  \tilde{\chi}_2^0 \to 3 l + E_{T}^{miss}$.
By contrast, in the NS scenario of the MSSM the leptons are very soft and hardly detectable due to the small mass splittings:
$\Delta_\pm, \Delta_0 \sim {\cal{O}} (1 {\rm GeV})$ \cite{NS-SUSY-Compressed}.

\item Since all the light particles in our scenario, i.e. $\tilde{\chi}_1^\pm$, $\tilde{\chi}_{1}^0$, $\tilde{\chi}_{2}^0$ and $\tilde{\chi}_{3}^0$,
have sizable $\tilde{H}_u$ component, the main decay modes of $\tilde{t}_1$ include
\begin{eqnarray}
\tilde{t}_1 \to b \tilde{\chi}_1^+ \to  b W^{\ast} \tilde{\chi}_1^0, \quad \tilde{t}_1 \to t \tilde{\chi}_{1}^0, \quad \tilde{t}_1 \to t \tilde{\chi}_{2,3}^0 \to  t X^{0 \ast} \tilde{\chi}_1^0,  \label{Higgsino-LSP}
\end{eqnarray}
and each mode corresponds to different signals. Because the signal of $\tilde{t}_1$ pair  production is shared by rich final states,
the LHC bounds on $m_{\tilde{t}_1}$ are usually weakened.
\end{itemize}
Moreover, we remind that the phenomenology of our scenario is different from that of the NS scenario with a Singlino-dominated DM. This can be seen for example
from the decay modes of $\tilde{t}_1$, which are presented in Eq.(\ref{Singlino-LSP}) and Eq.(\ref{Higgsino-LSP}) respectively.
We also remind that our scenario was scarcely discussed in literatures. In fact, within our knowledge only the work \cite{SNMSSM-Higgsino}
briefly commented that $\tilde{\chi}_1^0$ may be Higgsino-dominated in the constrained NMSSM.

This work is organized as follows. In Sec. II, we briefly recapitulate the framework of the NMSSM, then we scan its parameter space
by considering various constraints to get the NS scenarios in the NMSSM.  Especially, we take great pains to implement the constraints from the LHC
searches for SUSY by multiple packages and also by detailed Monte Carlo simulations, like what the work \cite{Natural-NMSSM-Simulation} did.
After these preparation, we exhibit in Sec. III the features of the NS scenario with a Higgsino-dominated DM, including its favored
spectrum and the properties of the neutralinos and stops, and subsequently in Sec. IV we take several benchmark points as examples to
show the detection of our scenario in future experiments. Finally, we draw our conclusions in Sec.V.  The details of our treatment on
the LHC searches for SUSY are presented in the Appendix.

\section{The Structure of the NMSSM and Our Scan Strategy}


\subsection{The Structure of the NMSSM}

The NMSSM extends the MSSM by adding one gauge singlet superfield $\hat{S}$, and since it aims at solving the $\mu$ problem of the MSSM,
a $Z_3$ discrete symmetry under which the Higgs superfields $\hat{H}_{u,d}$ and $\hat{S}$ are charged is adopted
to avoid the appearance of dimensionful parameters in its superpotential. Consequently, the superpotential of the NMSSM can be written as \cite{NMSSM-1}
\begin{eqnarray}
  W_{\rm NMSSM} &=& W_F + \lambda\hat{H_u} \cdot \hat{H_d} \hat{S}
  +\frac{1}{3}\kappa \hat{S^3},
 \end{eqnarray}
where $W_F$ is the superpotential of the MSSM without the $\mu$-term, and the dimensionless parameters $\lambda$, $\kappa$
describe the interactions among the Higgs superfields.

The Higgs potential of the NMSSM is given by the usual F-term and
D-term of the superfields as well as the soft breaking terms, which are given by
\begin{eqnarray}
V_{\rm NMSSM}^{\rm soft} &=& m_{H_u}^2 |H_u|^2 + m_{H_d}^2|H_d|^2
  + m_S^2|S|^2 +( \lambda A_{\lambda} SH_u\cdot H_d
  +\frac{1}{3}\kappa A_{\kappa} S^3 + h.c.), \label{input-parameter1}
\end{eqnarray}
with $H_u$, $H_d$ and $S$ representing the scalar component fields of $\hat{H}_u$,  $\hat{H}_d$ and $\hat{S}$
respectively. Considering that the physical implication of the fields $H_u$ and $H_d$ is less clear,
one usually introduces following combinations \cite{NMSSM-1}
\begin{eqnarray}
H_1=\cos\beta H_u + \varepsilon \sin\beta H_d^*, ~~
H_2=\sin\beta H_u - \varepsilon \cos\beta H_d^*, ~~H_3 = S,
\end{eqnarray}
where $\varepsilon$ is second-order antisymmetric tensor with $\varepsilon_{12}=-\varepsilon_{21}=1$ and $\varepsilon_{11}=\varepsilon_{22}=0$,
 and $\tan \beta \equiv v_u/v_d$ with $v_u$ and $v_d$ denoting
the vacuum expectation values of $H_u$ and $H_d$ respectively. In this representation, the redefined fields $H_i$ ($i=1,2,3$) are given by
\begin{eqnarray}
H_1 = \left ( \begin{array}{c} H^+ \\
       \frac{S_1 + i P_1}{\sqrt{2}}
        \end{array} \right),~~
H_2 & =& \left ( \begin{array}{c} G^+
            \\ v + \frac{ S_2 + i G^0}{\sqrt{2}}
            \end{array} \right),~~
H_3  = v_s +\frac{1}{\sqrt{2}} \left(  S_3 + i P_2 \right).
\label{fields}
\end{eqnarray}
These expressions indicate that the field $H_2$ corresponds to the SM Higgs doublet with $G^+$ and $G^0$
denoting the Goldston bosons eaten by $W$ and $Z$ bosons respectively, and the field $H_1$ represents a new $SU(2)_L$ doublet
scalar field which has no tree-level couplings to the W/Z bosons. These expressions also indicate that the Higgs sector of the
NMSSM includes three CP-even mass eigenstates $h_1$, $h_2$ and $h_3$ which are the
mixtures of the fields $S_1$, $S_2$ and $S_3$, two CP-odd mass eigenstates $A_1$ and $A_2$ which are composed by the fields $P_1$ and $P_2$,
as well as one charged Higgs $H^+$. In the following, we assume $m_{h_3} > m_{h_2} > m_{h_1}$ and $m_{A_2} > m_{A_1}$, and
call the state $h_i$ the SM-like Higgs boson if its dominant component comes from the field $S_2$.

In the NMSSM, the squared mass of the filed $S_2$ is given by
\begin{eqnarray}
m_{S_2 S_2}^2 = m_Z^2 \cos^2 2 \beta + \lambda^2 v^2  \sin^2 2 \beta, \nonumber
\end{eqnarray}
where the last term on the right side is peculiar to any singlet extension of the MSSM\cite{NMSSM-1},
and its effect is to enhance the mass of the SM-like Higgs boson in comparison with the case of the MSSM.
Moreover, if the inequation $m_{S_3 S_3}^2 < m_{S_2 S_2}^2$ holds, the mixing of the field $S_2$ with the
field $S_3$ in forming the SM-like Higgs boson can further enhance the mass. In
this case, $h_1$ is a singlet-dominate scalar, $h_2$ acts as the SM-like Higgs boson, and due to the enhancement
effects the requirement $m_{h_2} \simeq 125 {\rm GeV}$ does not necessarily need the large radiative correction of
stops \cite{NMSSM-Early-1,NMSSM-Early-4}. We remind that the singlet-dominated physical scalars (i.e. the mass eigenstates
mainly composed by $S_3$ and $P_2$ respectively)
are experimentally less constrained, and in case that they are lighter than about $200 {\rm GeV}$, $\mu = \lambda v_s$ naturally lies within
the range from $100 {\rm GeV}$ to $200 {\rm GeV}$.

In practice, it is convenient to use \cite{NMSSM-1}
\begin{eqnarray}
\lambda, \quad \kappa, \quad \tan \beta, \quad \mu, \quad M_A, \quad A_\kappa,  \label{input-parameter}
\end{eqnarray}
as input parameters, where $m_{H_u}^2$, $m_{H_d}^2$ and $m_S^2$ in Eq.(\ref{input-parameter1}) are traded
for $m_Z$, $\tan \beta$ and $\mu $ by the potential minimization conditions, and
$A_\lambda$ is replaced by the squared mass of the CP-odd field $P_1$, which is given by
\begin{eqnarray}
M^2_A \equiv m_{P_1 P_1}^2 = \frac{2\mu}{\sin2\beta}(A_{\lambda}+\kappa v_s).
\end{eqnarray}
Note that $M_A$ represents the mass scale of the new doublet $H_1$, and it is preferred by current experiments to
be larger than about $300 {\rm GeV}$.

When we discuss the naturalness of the NMSSM, we consider two fine tuning quantities defined by \cite{Baer:2013gva}
\begin{eqnarray}
\Delta_Z = \max_i |\frac{\partial \log m_Z^2}{\partial \log p_i}|, \quad \Delta_h = \max_i |\frac{\partial \log m_h^2}{\partial \log p_i}|,
\end{eqnarray}
where $h$ represents the SM-like Higgs boson, $p_i$ denotes SUSY parameters at the weak scale, and it includes the parameters listed in Eq.(\ref{input-parameter}) and
top quark Yukawa coupling $Y_t$ with the latter used to estimate the sensitivity to stop masses. Obviously,
$\Delta_Z$ ($\Delta_h$) measures the sensitivity of $m_Z$ ($m_h$) to SUSY parameters at weak scale,
and the larger its value becomes, the more tuning is needed to get the corresponding mass \footnote{As was pointed out in \cite{Baer:2013gva},
if the NMSSM is considered as the low energy realization of an (unknown) overarching ultimate theory,
$\Delta_Z$ and $\Delta_h$ can be thought of as providing a lower bound
on electroweak fine-tuning. Any parameter point with  a low $\Delta_Z$ and $\Delta_h$  implies that
the ultimate theory may be low fine-tuned at high energy scale. By contrast, if the point correspond to large
$\Delta_Z$ and $\Delta_h$, the underlying theory must be fine-tuned.}. In our calculation, we calculate
 $\Delta_Z$ and  $\Delta_h$ by the formulae presented in \cite{Ulrich-Fine-Tuning} and \cite{lambd-SUSY-recent-2} respectively.

The NMSSM predicts five neutralinos, which are the mixtures of the fields Bino $\tilde{B}^0$, Wino $\tilde{W}^0$, Higgsinos $\tilde{H}_{d,u}^0$ and
Singlino $\tilde{S}^0$.  In the basis $\psi^0 = (-i \tilde{B}^0, - i \tilde{W}^0, \tilde{H}_{d}^0, \tilde{H}_{u}^0, \tilde{S}^0)$, the neutralino mass
matrix is given by \cite{NMSSM-1}
\begin{equation}
{\cal M} = \left(
\begin{array}{ccccc}
M_1 & 0 & -\frac{g_1 v_d}{\sqrt{2}} & \frac{g_1 v_u}{\sqrt{2}} & 0 \\
  & M_2 & \frac{g_2 v_d}{\sqrt{2}} & - \frac{g_2 v_u}{\sqrt{2}} &0 \\
& & 0 & -\mu & -\lambda v_u \\
& & & 0 & -\lambda v_d\\
& & & & \frac{2 \kappa}{\lambda} \mu
\end{array}
\right). \label{eq:MN}
\end{equation}
If $|M_1|, |M_2| \gg |\mu|$ and $2 \kappa/\lambda \sim 1$, the Bino and Wino fields are decoupled from the rest fields. In this case, the remaining three light
neutralinos $\tilde{\chi}_i^0$ ($i=1,2,3$) can be approximated by
\begin{eqnarray}
\tilde{\chi}_i^0 & \approx & N_{i3} \tilde{H}_d^0 + N_{i4}\tilde{H}_u^0 + N_{i5} \tilde{S}^0, \label{neutralino-mass}
\end{eqnarray}
where the elements of the rotation matrix $N$ roughly satisfy
\begin{eqnarray}
N_{i3}:N_{i4}:N_{i5} \simeq \lambda (v_d \mu -  v_u m_{\tilde{\chi}_i^0}):
\lambda (v_u \mu -  v_d m_{\tilde{\chi}_i^0}): (m_{\tilde{\chi}_i^0}^2 - \mu^2)  \label{neutralino-mixing}
\end{eqnarray}
with $m_{\tilde{\chi}_i^0}$ denoting the mass of $\tilde{\chi}_i^0$. In the following, we are interested in the
parameter region featured by $\tan \beta \sim 10$, $2 \kappa/\lambda \sim (1 - 1.5)$ and
$|\mu| \sim (100-200) {\rm GeV}$, which is hereafter dubbed as the NS scenario with $\tilde{\chi}_1^0$ being Higgsino-dominated.
From Eq.(\ref{neutralino-mass})
and Eq.(\ref{neutralino-mixing}), one can conclude that this scenario has following characters
\begin{itemize}
\item The lightest two neutralinos $\tilde{\chi}_1^0$ and $\tilde{\chi}_2^0$ are Higgsino-dominated, and
$\tilde{\chi}_3^0$ is Singlino-dominated. Their masses should satisfy following relations: $m_{\tilde{\chi}_1^0} < |\mu|$, $m_{\tilde{\chi}_2^0} \sim |\mu|$
and  $m_{\tilde{\chi}_3^0} > |\frac{2 \kappa}{\lambda} \mu| >  |\mu| $.
\item As far as $\tilde{\chi}_1^0$ is concerned, its largest component comes from $\tilde{H}_{u}^0$ field. If the splitting between
$m_{\tilde{\chi}_1^0}$ and $|\mu|$ is significant, its Singlino component may also be quite large. The importance of the Singlino component
is that it can dilute the couplings of $ \tilde{\chi}_1^0$ with $W$ and $Z$ bosons, Higgs scalars and SM fermions, and consequently the density of
$ \tilde{\chi}_1^0$ can coincide with the DM density measured by WMAP and Planck experiments.
\item The $\tilde{H}_u^0$ and $\tilde{H}_d^0$ components in $ \tilde{\chi}_2^0$ should be comparable, and they are usually much larger than $\tilde{S}^0$
component of  $ \tilde{\chi}_2^0$, i.e. $|N_{23}| \sim |N_{24}| \gg |N_{25}|$.
\item As for $ \tilde{\chi}_3^0$, the relation $|N_{35}| > |N_{33}| >  |N_{34}|$ usually holds.
\end{itemize}

\subsection{Strategy in Scanning the Parameter Space of the NMSSM}

In this part, we perform a comprehensive scan over the parameter space of the NMSSM by considering various experimental constraints.
Especially, we take great pains to implement the constraints from the direct searches for SUSY at the LHC. After this procedure, we
get the NS scenario with a Higgsino-dominated $\tilde{\chi}_1^0$.

\begin{table}[t]
\caption{Favored parameter ranges for different types of samples. In each item, the
range in the first row is for the samples that survive the constraints (1), (2) and (3) presented in the text,
and that in the second row corresponds to the samples that further satisfy the constraints from the direct search for
sparticles at the LHC Run-I, i.e. the constraints (4) and (5). The quantity $R$ in the last row represents the retaining ratio
of the samples before and after considering the direct search constraints in our scan.  For Type I samples,
$h_1$ corresponds to the SM-like Higgs boson and $\tilde{\chi}_1^0$ is Bino-dominated,
while for Type II, III and IV samples, $h_2$ acts as the SM-like Higgs boson with $\tilde{\chi}_1^0$
being Bino-, Singlino- and Higgsino-dominated respectively.\label{table-1} } \centering

\vspace{0.2cm}

\begin{tabular}{|c|c|c|c|c|}
\cline{1-5}
\multicolumn{1}{|c|}{\multirow{1}{*}{}}
& \multicolumn{1}{c|}{Type~I  }
& \multicolumn{1}{c|}{Type~II  }
& \multicolumn{1}{c|}{Type~III  }
&  \multicolumn{1}{c|}{Type~IV  } \\
\cline{1-5}
\multicolumn{1}{|c|}{\multirow{2}{*}{~$\lambda$~}}
& $0.\thicksim0.41$ & $0.\thicksim0.68$
& $0.\thicksim0.75$ & $0.18\thicksim 0.71$ \\
\multicolumn{1}{|c|}{}
& $0.\thicksim0.41$& $0.\thicksim0.68$
& $0.\thicksim0.75$& $0.18\thicksim 0.71$\\
\cline{1-5}
\multicolumn{1}{|c|}{\multirow{2}{*}{~~~$\kappa$~~~}}
& $ 0.\thicksim0.66 $ & $ 0.\thicksim0.52 $
& $ 0.\thicksim0.27 $ & $ 0.\thicksim0.51 $ \\
\multicolumn{1}{|c|}{}
& $ 0.\thicksim0.66 $ & $ 0.\thicksim0.52 $
& $ 0.\thicksim0.27 $ & $ 0.\thicksim0.51 $ \\
\cline{1-5}
\multicolumn{1}{|c|}{\multirow{2}{*}{~~~$\tan\beta$~~~}}
& $ 3\thicksim 60$ & $ 4\thicksim 38 $
& $ 3\thicksim 60$ & $ 3\thicksim 18 $  \\
\multicolumn{1}{|c|}{}
& $ 3\thicksim 60 $ & $ 4\thicksim 38 $
& $ 3\thicksim 60$ & $ 3\thicksim 18 $  \\
\cline{1-5}
\multicolumn{1}{|c|}{\multirow{2}{*}{~~~$\mu$(GeV)}}
& $ 150\thicksim 400 $ & $ 115\thicksim 370 $
& $ 105\thicksim 315 $ & $ 110 \thicksim 175 $\\
\multicolumn{1}{|c|}{}
& $ 180 \thicksim 400 $ & $ 115\thicksim 370 $
& $ 105\thicksim 315 $ & $ 110 \thicksim 165 $  \\
\cline{1-5}
\multicolumn{1}{|c|}{\multirow{2}{*}{~~~$A_\kappa$(GeV)}}
& $ -2000\thicksim 0$ & $ -750\thicksim 0 $
& $ -350 \thicksim 40 $ & $ -650\thicksim -20 $  \\
\multicolumn{1}{|c|}{}
& $ -2000\thicksim 0 $ & $ -750 \thicksim 0 $
& $ -350\thicksim 40$ & $ -650\thicksim -30 $ \\
\cline{1-5}
\multicolumn{1}{|c|}{\multirow{2}{*}{$M_{Q_3}$(GeV)}}
& $ 390 \thicksim 2000 $ & $ 550 \thicksim 2000 $
& $ 450 \thicksim 2000 $ & $ 450 \thicksim 2000  $  \\
\multicolumn{1}{|c|}{}
& $ 570 \thicksim 2000 $ & $ 680 \thicksim 2000 $
& $ 450 \thicksim 2000 $ & $ 620 \thicksim 2000 $  \\
\cline{1-5}
\multicolumn{1}{|c|}{\multirow{2}{*}{$M_{U_3}$(GeV)}}
& $ 480 \thicksim 2000 $ & $ 530 \thicksim 2000 $
& $ 480 \thicksim 2000 $ & $490 \thicksim 2000 $ \\
\multicolumn{1}{|c|}{}
& $ 610 \thicksim 2000 $ & $ 680\thicksim 2000 $
& $ 580 \thicksim 2000 $ & $ 560 \thicksim 2000$ \\
\cline{1-5}
\multicolumn{1}{|c|}{\multirow{2}{*}{$M_{\tilde{l}}$(GeV)}}
& $ 100 \thicksim 1000 $ & $ 100 \thicksim 640 $
& $ 100 \thicksim 730 $ & $ 100 \thicksim 950 $ \\
\multicolumn{1}{|c|}{}
& $ 100 \thicksim 1000 $ & $ 100 \thicksim 640 $
& $ 100 \thicksim 730 $ & $ 100 \thicksim 950 $  \\
\cline{1-5}
\multicolumn{1}{|c|}{\multirow{2}{*}{$A_{t}$(GeV)}}
& $ -4700 \thicksim 5000 $ & $ -4500 \thicksim 4500 $
& $ -5000 \thicksim 4800 $ & $ -5000\thicksim 4500 $ \\
\multicolumn{1}{|c|}{}
& $ -4400 \thicksim 4850 $ & $ -4500 \thicksim 4500 $
& $ - 5000 \thicksim 4600 $ & $ -5000 \thicksim 4500  $  \\
\cline{1-5}
\multicolumn{1}{|c|}{\multirow{2}{*}{~~~$M_1$(GeV)}}
& $ 40 \thicksim 350$ & $ 20 \thicksim 175 $
& $ 45 \thicksim 500 $ & $ 120 \thicksim 500 $\\
\multicolumn{1}{|c|}{}
& $ 40 \thicksim 350 $ & $ 40 \thicksim 175$
& $ 45 \thicksim 500 $ & $ 120 \thicksim 500 $ \\
\cline{1-5}
\multicolumn{1}{|c|}{\multirow{2}{*}{$M_2$(GeV)}}
& $ 105 \thicksim 1000 $ & $ 150 \thicksim 1000 $
& $ 150 \thicksim 1000$ & $ 160 \thicksim 1000 $  \\
\multicolumn{1}{|c|}{}
& $ 105 \thicksim 1000$ &  $ 150 \thicksim 1000 $
& $ 155 \thicksim 1000 $ &  $ 165 \thicksim 1000 $ \\
\cline{1-5}
\multicolumn{1}{|c|}{\multirow{2}{*}{~~~$M_A$(GeV)}}
& $ 200 \thicksim 2000 $ & $ 900 \thicksim 2000 $
& $ 500 \thicksim 2000 $ & $ 430 \thicksim 2000 $ \\
\multicolumn{1}{|c|}{}
& $ 320 \thicksim 2000 $ & $ 900 \thicksim 2000 $
& $ 500 \thicksim 2000 $ & $ 430 \thicksim 2000 $ \\
\hline
$R$& $52\%$ & $54\%$ & $71\%$ & $65\%$ \\
\hline
\end{tabular}
\end{table}

We begin our study by making following assumptions about some unimportant SUSY parameters:
\begin{itemize}
\item We fix all soft breaking parameters for the first two generation squarks
at $2 {\rm TeV}$. Considering that the third generation squarks can affect significantly
the mass of the SM-like Higgs boson, we vary freely all soft parameters in this sector except
that we assume $m_{U_3} = m_{D_3}$ for right-handed soft breaking masses and $A_t = A_b$
for soft breaking trilinear coefficients.
\item Considering that we require the NMSSM to explain the discrepancy of the measured value
of the muon anomalous magnetic moment from its SM prediction, we treat the common value for
all soft breaking parameters in the slepton sector (denoted by $m_{\tilde{l}}$ hereafter) as
a free parameter.
\item We fix gluino mass at $2 {\rm TeV}$, and treat the Bino mass $M_1$ and the Wino mass $M_2$
as free parameters since they affect the properties of the neutralinos.
\end{itemize}
Then we use the package NMSSMTools-4.9.0 \cite{NMSSMTools}
to scan following parameter space:
\begin{eqnarray}\label{NMSSM-scan}
&& 0 <\lambda\leq 0.75,\quad  0 <\kappa \leq 0.75, \quad  2 \leq \tan{\beta} \leq 60,\quad  100{\rm ~GeV}\leq m_{\tilde{l}} \leq 1 {\rm ~TeV},  \nonumber \\
&& 100 {\rm GeV} \leq \mu \leq 1 {\rm TeV}, \quad 50 {\rm ~GeV}\leq M_A \leq 2 {\rm ~TeV}, \quad |A_{\kappa}| \leq 2 {\rm TeV}, \nonumber\\
&& 100{\rm ~GeV}\leq M_{Q_3},M_{U_3} \leq 2 {\rm ~TeV}, \quad  |A_{t}|\leq {\rm min}(3 \sqrt{M_{Q_3}^2 + M_{U_3}^2}, 5 {\rm TeV}), \nonumber\\
&& 20 {\rm GeV} \leq M_1 \leq 500 {\rm GeV}, \quad 100 {\rm GeV} \leq M_2 \leq 1 {\rm TeV},
\end{eqnarray}
where all the parameters are defined at the scale of $1 {\rm TeV}$.
During the scan, we use following constraints to select physical parameter points:
\begin{itemize}
\item[(1)] All the constraints implemented in the package NMSSMTools-4.9.0, which include
the $Z$-boson invisible decay, the LEP search for sparticles (i.e. the lower bounds on various
sparticle masses and the upper bounds on the chargino/neutralino pair production rates),
the $B$-physics observables such as the branching ratios for $B \to X_s \gamma$ and
$B_s \to \mu^+ \mu^-$, the discrepancy of the muon anomalous magnetic moment, the dark matter relic density and the LUX
limits on the scattering rate of dark matter with nucleon. In getting the constraint from a certain
observable which has an experimental central value, we use its latest measured result and require
the NMSSM to explain the result at $2\sigma$ level.

\item[(2)] Constraints from the direct searches for Higgs bosons at LEP, Tevatron and LHC.
Especially we require that one CP-even Higgs boson acts as the SM-like Higgs boson discovered at the LHC.
We implement these constraints with the packages HiggsSignal for $125 {\rm GeV}$ Higgs data fit
\cite{HiggsSignals} \footnote{In our fit, we adopt a moderately wider range of the SM-like Higgs boson mass,
$122 {\rm GeV} \leq m_h \leq 128 {\rm GeV}$, in comparison with the default uncertainty of $2 {\rm GeV}$ for $m_h$ in the
package HiggsSignal. This is because $\lambda \gtrsim 0.4$  may induce a ${\cal{O}}(1 {\rm GeV})$ correction to $m_h$
at two-loop level \cite{Goodsell:2014pla,Staub:2015aea}, which is not considered in the NMSSMTools.}
and HiggsBounds for non-standard Higgs boson search at colliders \cite{HiggsBounds}.

\item[(3)] Constraints from the fine-tuning consideration: $\Delta_Z \leq 50$ and $\Delta_h \leq 50$.

\item[(4)] Constraints from the preliminary analyses of the ATLAS and CMS groups in their direct
searches for sparticles at the LHC Run-I. We implement these constraints by the packages FastLim \cite{Papucci:2014rja}
and SModelS \cite{Kraml:2013mwa}.  These two packages provide cut efficiencies or upper bounds on some
sparticle production processes in simplified model framework, and thus enable us to impose the direct search
bounds in an easy and fast way.  In the appendix, we briefly introduce the two packages.

\item[(5)] Constraints from the latest searches for electroweakinos and stops by the ATLAS
collaboration at the LHC Run-I. We implement these constraints by detailed Monte Carlo simulation. Since we have to treat more than
twenty thousand samples at this step, this process is rather time consuming in our calculation by clusters. In the appendix,
we provide details of our simulation.
\end{itemize}

\begin{figure}[t]
\centering
\includegraphics[height=8cm,width=16cm]{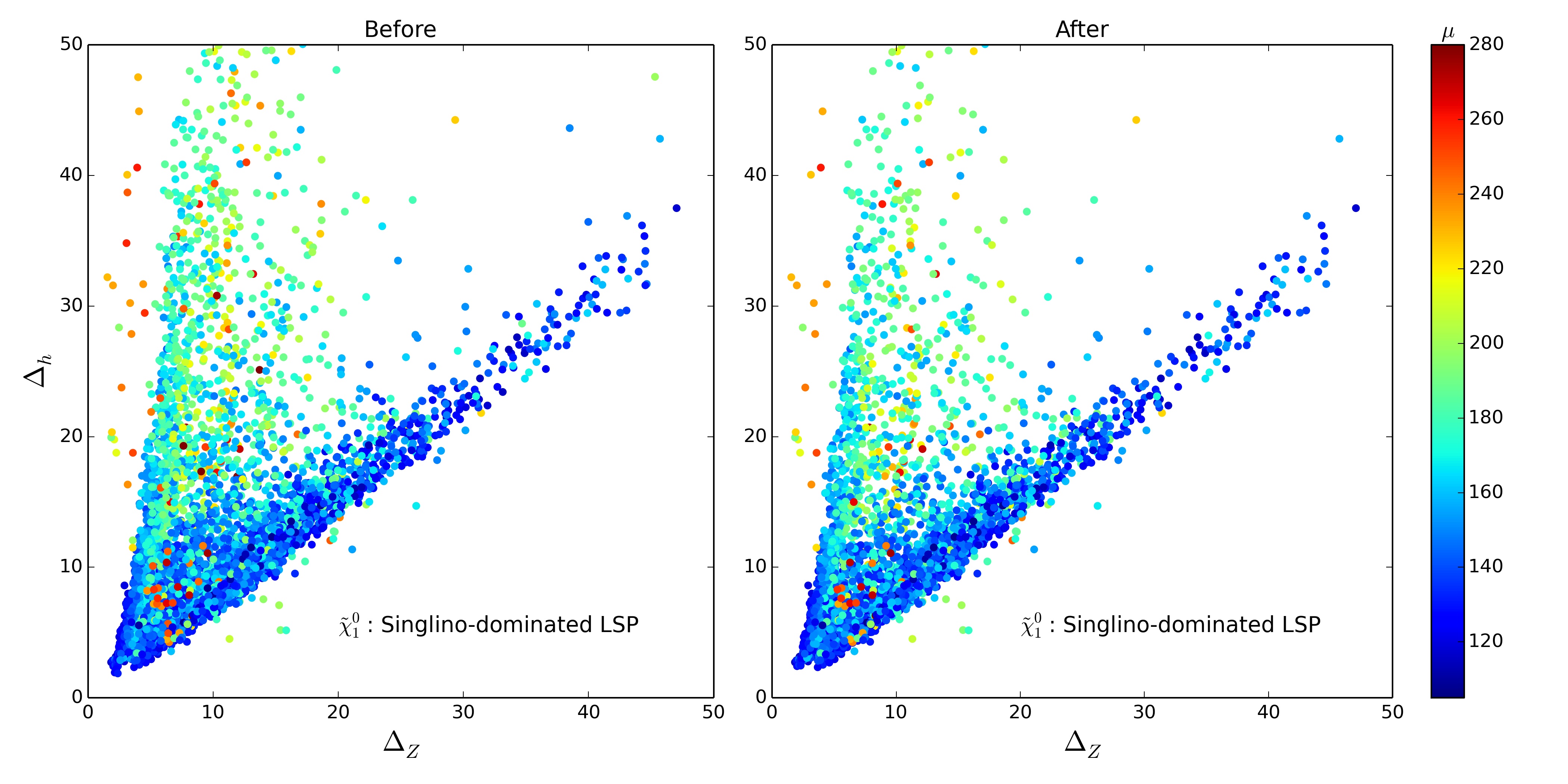} \\

\vspace{-0.3cm}

\includegraphics[height=8cm,width=16cm]{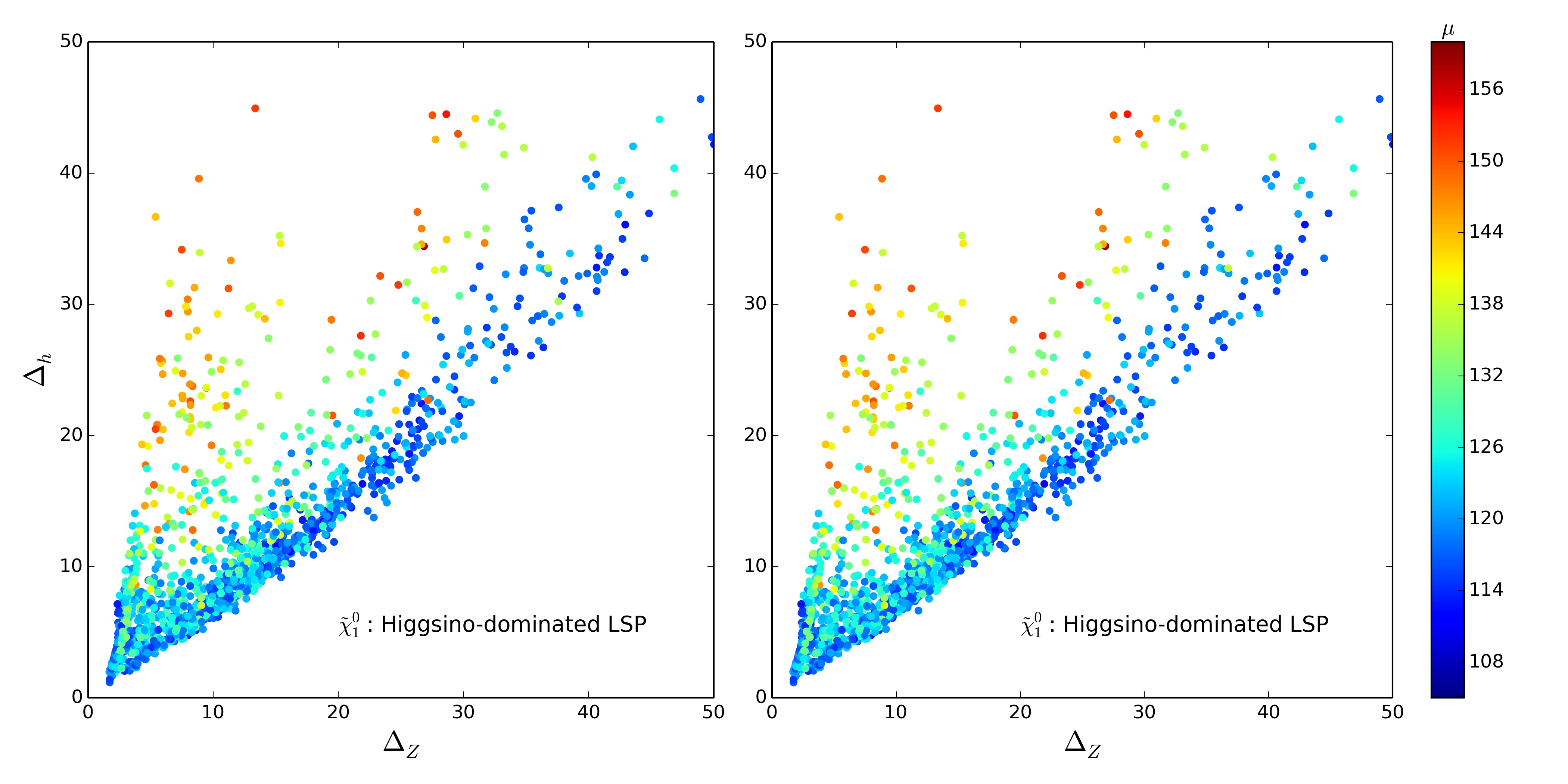}
\vspace{-1cm}
\caption{$\Delta_Z$ and $\Delta_h$ predicted by the Type III samples (upper panels) and Type IV samples (lower panels) respectively.
Samples in the left panels survive the constraints (1), (2) and (3) presented in the text, and those in the right panels further
satisfy the constraints from the direct searches for sparticles at LHC Run I, i.e. the constraints (4) and (5). The panels in each
row adopt same color convention for $\mu$, which is presented on the right side of the row.  \label{fig1}}
\end{figure}

After analyzing the samples that survive the constraints, we find that they can be classified into four types:
for Type I samples, $h_1$ corresponds to the SM-like Higgs boson and $\tilde{\chi}_1^0$ is Bino-dominated,
while for Type II, III and IV samples, $h_2$ acts as the SM-like Higgs boson with $\tilde{\chi}_1^0$
being Bino-, Singlino- and Higgsino-dominated respectively.  In Table \ref{table-1}, we list the favored parameter ranges for each type of samples
before and after considering the constraints from the direct search experiments, i.e. the constraints (4) and (5). Note that in the last row
of this table, the retaining ratio $R$ is defined by $R = N_{after}/N_{before}$ where $N_{before}$ is the number of the samples that satisfy
the constraints (1), (2) and (3) in our scan, and $N_{after}$ is the number of the samples that further satisfy the constraints (4) and (5).
This table indicates that although the direct search experiments are effective in excluding parameter points encountered in the scan, they
scarcely change the ranges of the input parameters where $\Delta_Z$ and $\Delta_h$ take rather low values, or equivalently speaking the NMSSM can
naturally predict $m_Z$ and $m_h$ even after considering the direct search constraints from LHC Run-I. 
The underlying reason for this phenomenology is that the exclusion capability of the
direct search experiments depends not only on sparticle production rate, but also on the decay chain of the sparticle and the mass
gap between the sparticle and its decay product. We checked that a large portion of the excluded samples are characterized by $M_2 \leq 250 {\rm GeV}$.
In this case, there exist one moderately light neutralino and one moderately light chargino with both of them being Wino-dominated, and their
associated production rate at the LHC is quite large so that the $3 l + E_{T}^{miss}$ signal of the production
after cuts may exceed its experimental upper bound (see appendix for more information).  Among the four types of points, we also find that the lowest
fine-tuning comes from type III and type IV samples, for which $\Delta_Z$ and $\Delta_h$ may be as low as
about 2. This character is shown in Fig.\ref{fig1}, where we project type III and IV samples on $\Delta_Z-\Delta_h$
plane. We emphasize that for the type IV samples with $\Delta_Z, \Delta_h \lesssim 10$, $\mu$ is upper bounded by about $145 {\rm GeV}$.
In this case, it is compressed spectrum among the Higgsino-dominated particles $\tilde{\chi}_1^0$, $\tilde{\chi}_2^0$ and $\tilde{\chi}_1^\pm$
that helps the samples evade the direct search experiments.

Because the type IV samples were scarcely studied in previous literatures and also because they have similar phenomenology to that of the NS scenario in the MSSM,
we in the following focus on this type of samples.  In order to make the essential
features of the samples clear, we only consider those that  satisfy additionally the condition $M_1, M_2, m_{\tilde{l}} \geq 300 {\rm GeV}$.
Hereafter we call such samples collectively as the NS scenario with a Higgsino-dominated $\tilde{\chi}_1^0$. As we will show below,
$|\mu|$ in this scenario is upper bounded by about $160 {\rm GeV}$, so the condition is equivalent to $M_1, M_2, m_{\tilde{l}} \gtrsim 2 |\mu| $.
In this case, gauginos and sleptons affect little on the properties of the lightest three neutralino, and the lighter chargino.

\begin{table}[t]
\caption{Similar to Table \ref{table-1}, but for the NS scenario of the NMSSM with $\tilde{\chi}_1^0$ being Higgsino dominant. This scenario is defined by
Type IV samples, and it further requires the samples to satisfy $M_1, M_2, m_{\tilde{l}} \geq 300 {\rm GeV}$. Quantities with mass dimension are in
unit of ${\rm GeV}$. \label{table-2}}

\centering

\vspace{0.2cm}

\begin{tabular}{|c|c|c|c|c|c|}
\cline{1-6}
para & range & para & range & para & range \\
\cline{1-6} \multicolumn{1}{|c|}
{\multirow{2}{*} {~$\tan \beta $~}} & $7\thicksim 18$ & \multicolumn{1}{|c|}{\multirow{2}{*}{~$m_{\tilde{\chi}_1^0}$~}}
& $65\thicksim85$ & \multicolumn{1}{|c|}{\multirow{2}{*}{~$A_\kappa$~}}& $-400\thicksim -60$ \\
\multicolumn{1}{|c|}{}& $7\thicksim 18$& \multicolumn{1}{|c|}{}
&  $65\thicksim85$& \multicolumn{1}{|c|}{} & $-400\thicksim -60$ \\
\cline{1-6} \multicolumn{1}{|c|}{\multirow{2}{*}{~$\kappa$~}}
& $0.15\thicksim0.49$ & \multicolumn{1}{|c|}{\multirow{2}{*}{~$ m_{\tilde{\chi}_2^0}$~}}
& $125\thicksim195$ & \multicolumn{1}{|c|}{\multirow{2}{*}{~$m_{h_1}$~}} & $45\thicksim 120 $ \\
\multicolumn{1}{|c|}{}& $0.15\thicksim0.49$& \multicolumn{1}{|c|}{}& $125\thicksim195$
& \multicolumn{1}{|c|}{}& $45\thicksim 120 $ \\
\cline{1-6}\multicolumn{1}{|c|}{\multirow{2}{*}{~$\lambda$~}}
& $0.28\thicksim 0.68$& \multicolumn{1}{|c|}{\multirow{2}{*}{~$m_{\tilde \chi_3^0}$~}}
& $150 \thicksim260 $& \multicolumn{1}{|c|}{\multirow{2}{*}{~$m_{A_1}$~}}
&   $120\thicksim 350$  \\
\multicolumn{1}{|c|}{}& $0.28\thicksim 0.68$& \multicolumn{1}{|c|}{} & $150 \thicksim260$
& \multicolumn{1}{|c|}{}&  $120\thicksim 350$ \\
\cline{1-6}\multicolumn{1}{|c|}{\multirow{2}{*}{~$\mu$~}}& $110 \thicksim 160 $
& \multicolumn{1}{|c|}{\multirow{2}{*}{~$m_{\tilde \chi_1^{\pm}}$~}}
& $105 \thicksim 150 $& \multicolumn{1}{|c|}{\multirow{2}{*}{~$m_{H^{\pm}}$~}}
& $800 \thicksim 2000 $ \\
\multicolumn{1}{|c|}{}
& $110 \thicksim 160 $& \multicolumn{1}{|c|}{}& $105\thicksim 150$
& \multicolumn{1}{|c|}{}& $800 \thicksim 2000 $\\
\cline{1-6}
\multicolumn{1}{|c|}{\multirow{2}{*}{~$A_t$~}}& $-5000\thicksim 4500 $
& \multicolumn{1}{|c|}{\multirow{2}{*}{~$m_{\tilde t_1}$~}} & $380 \thicksim 2050 $
& \multicolumn{1}{|c|}{\multirow{2}{*}{~$m_{\tilde t_2}$~}} &  $1050 \thicksim 3100 $ \\
\multicolumn{1}{|c|}{}& $-5000\thicksim 4500 $ & \multicolumn{1}{|c|}{}
& $500 \thicksim 2050 $ & \multicolumn{1}{|c|}{} &  $1100 \thicksim 3100 $ \\
\cline{1-6}
\end{tabular}
\end{table}

\section{Key features of the NS scenario with $\tilde{\chi}_1^0$ being Higgsino-dominated}

In this section, we investigate the features of the NS scenario with $\tilde{\chi}_1^0$ being Higgsino-dominated. We are particulary interested
in neutralino-chargino sector and stop sector since they play an important role in determining the fine tunings of the theory.
In Table \ref{table-2}, we show the favored ranges of some quantities such as $\mu$ and stop masses. This table indicates that
our scenario is featured by $\mu/{\rm GeV} \in [ 110, 160 ]$,  $m_{\tilde{\chi}_1^0}/{\rm GeV} \in [ 65, 85 ]$,
$ m_{\tilde{\chi}_2^0}/{\rm GeV} \in [ 125, 195 ]$, $ m_{\tilde{\chi}_3^0}/{\rm GeV} \in [ 150, 260 ]$,
$m_{\tilde{\chi}_1^\pm}/{\rm GeV} \in [105, 150 ]$ and $m_{\tilde{t}_1}/{\rm GeV} \in [ 500,  2050 ]$, and among the ranges,
only the lower bound of $m_{\tilde{t}_1}$ is shifted from $380 {\rm GeV}$ to $500 {\rm GeV}$ by the constraints from the direct
search experiments. Moreover, we checked that in our scenario, the ratio $2 \kappa/\lambda$ is restricted in the range from about
1 to 1.5. In this case, Higgsinos and Singlino are approximately degenerated in mass, and consequently they mix strongly to form mass
eigenstates.

\begin{figure}[t]
\centering
\includegraphics[height=6cm,width=5.2cm]{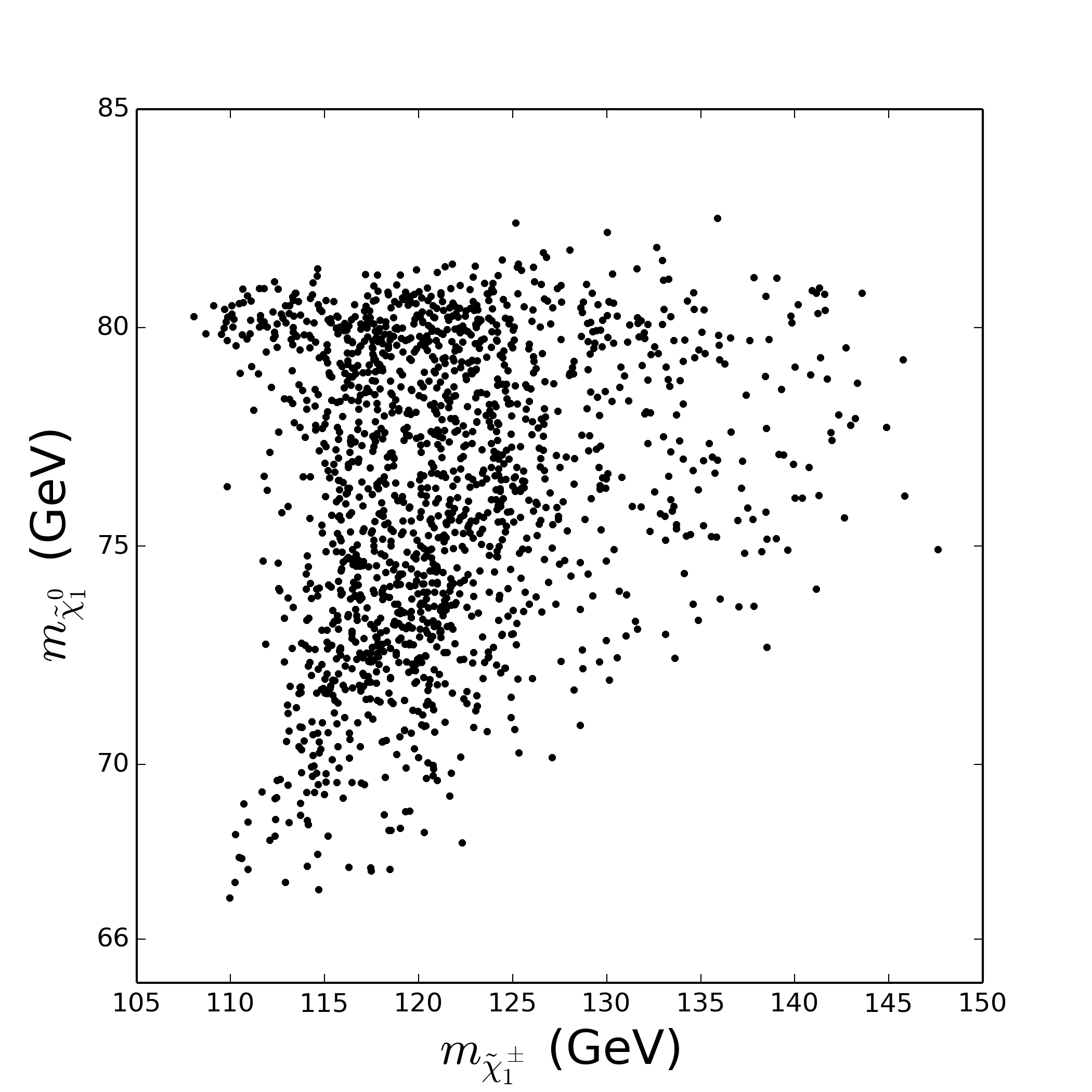} \hspace{-0.6cm}
\includegraphics[height=6cm,width=5.2cm]{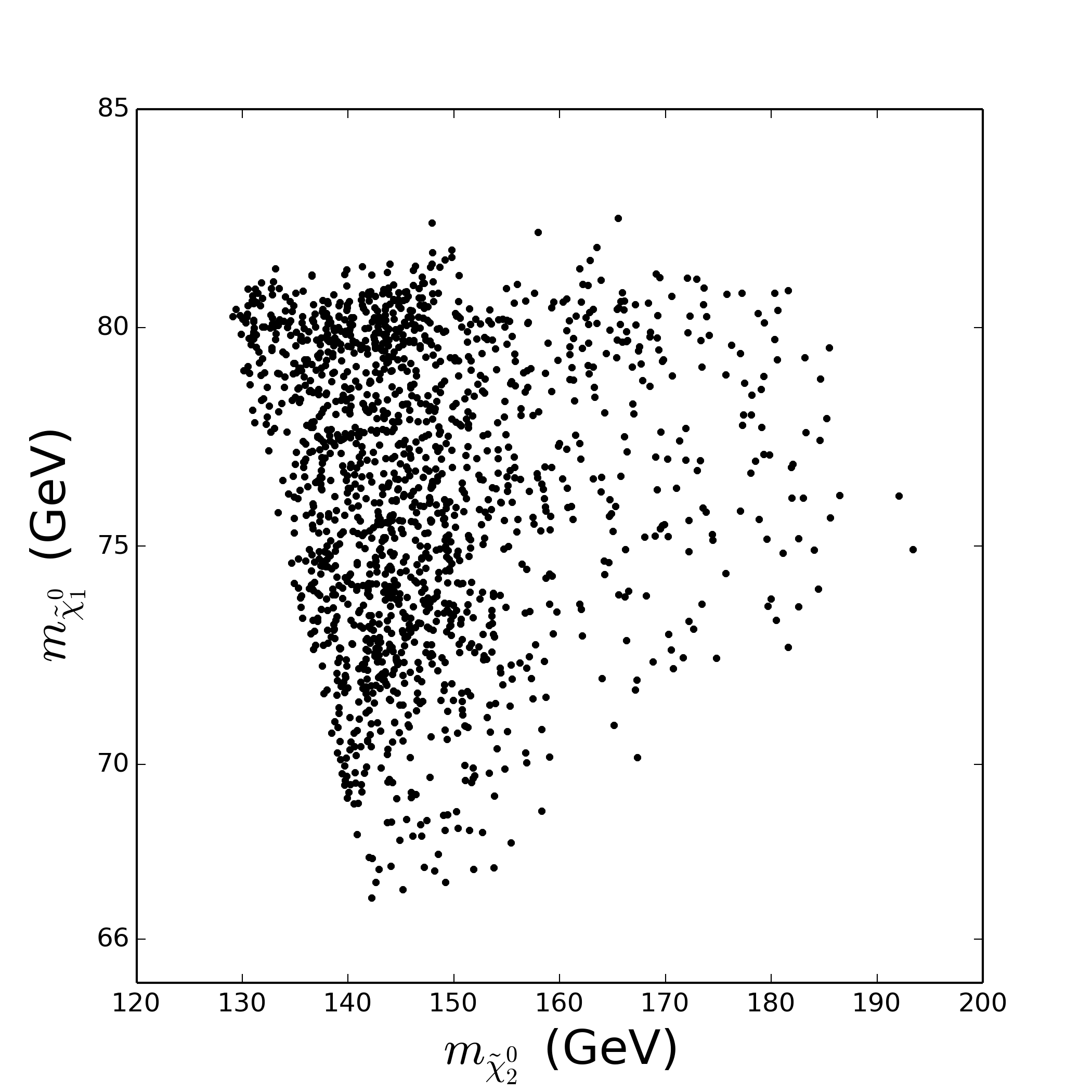}  \hspace{-0.6cm}
\includegraphics[height=6cm,width=5.2cm]{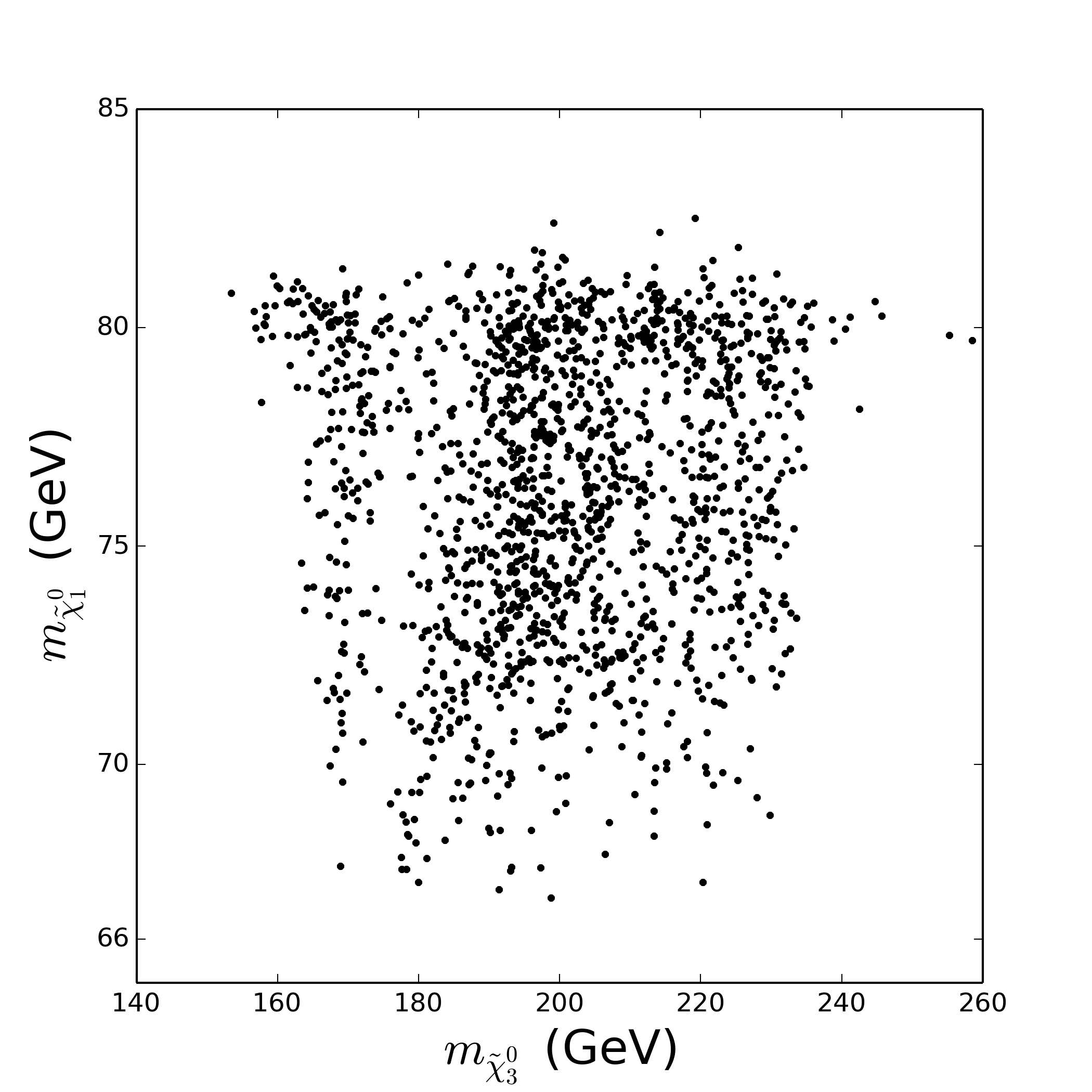}
\vspace{-0.4cm}
\caption{Mass spectrum of $\tilde{\chi}_1^0$, $\tilde{\chi}_2^0$, $\tilde{\chi}_3^0$ and $\tilde{\chi}_1^\pm$ in the NS scenario
with $\tilde{\chi}_1^0$ being Higgsino-dominated. Note that the mass splittings among these particles are induced by
the strong mixings between Higgsinos and Singlino, and significantly larger than those in the NS scenario of the MSSM.
\label{fig2}}
\end{figure}

\begin{figure}[t]
\centering
\includegraphics[height=6cm,width=5.2cm]{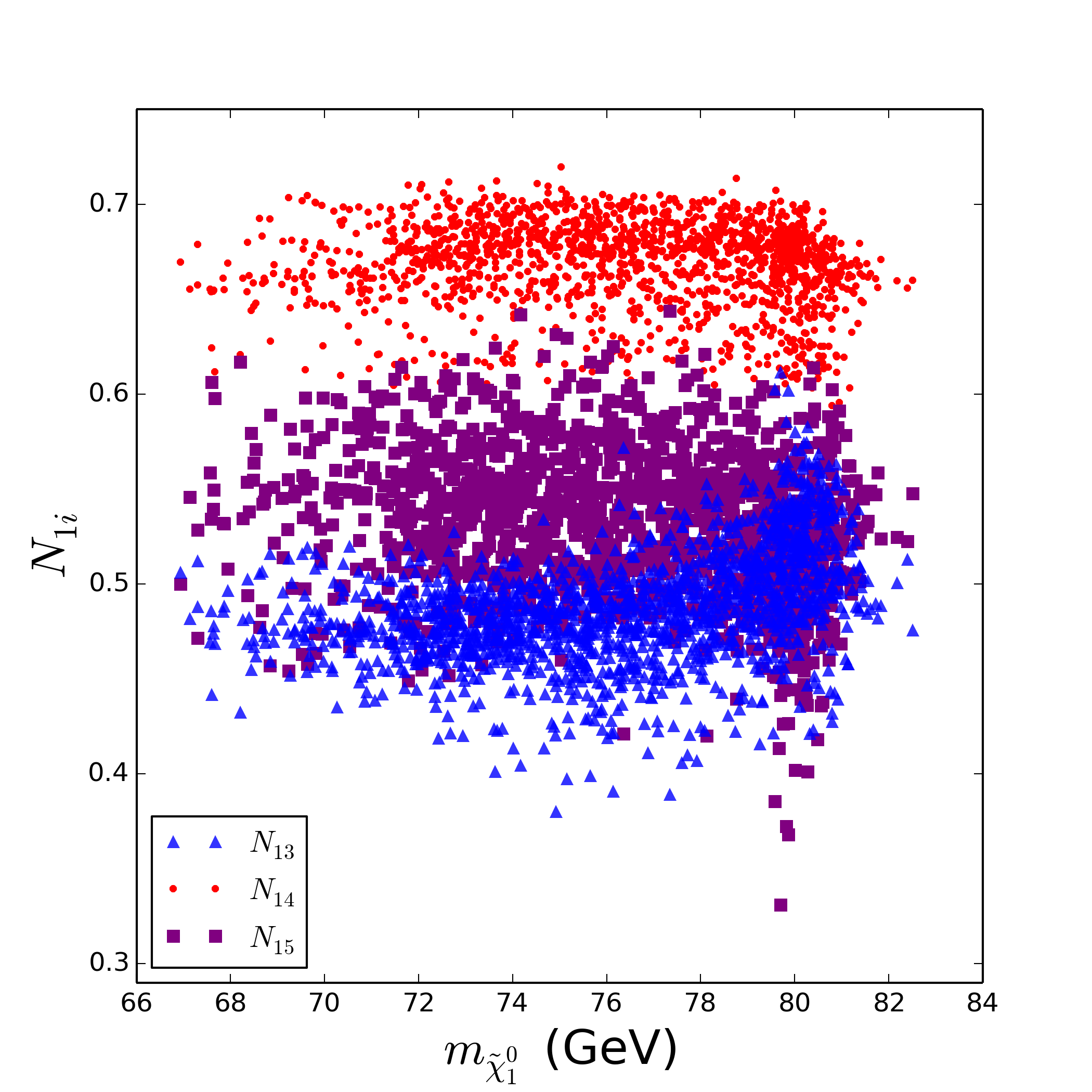} \hspace{-0.6cm}
\includegraphics[height=6cm,width=5.2cm]{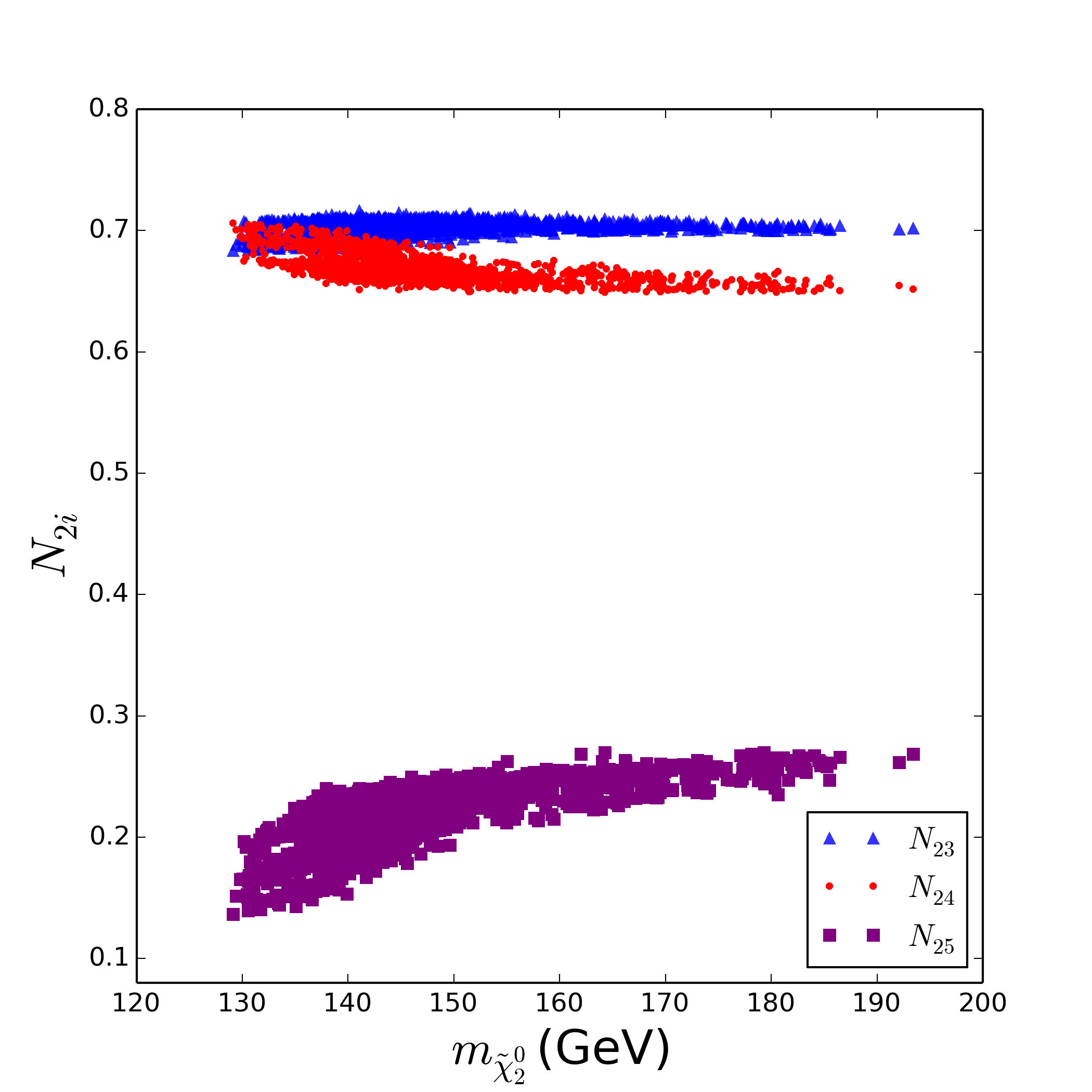}  \hspace{-0.6cm}
\includegraphics[height=6cm,width=5.2cm]{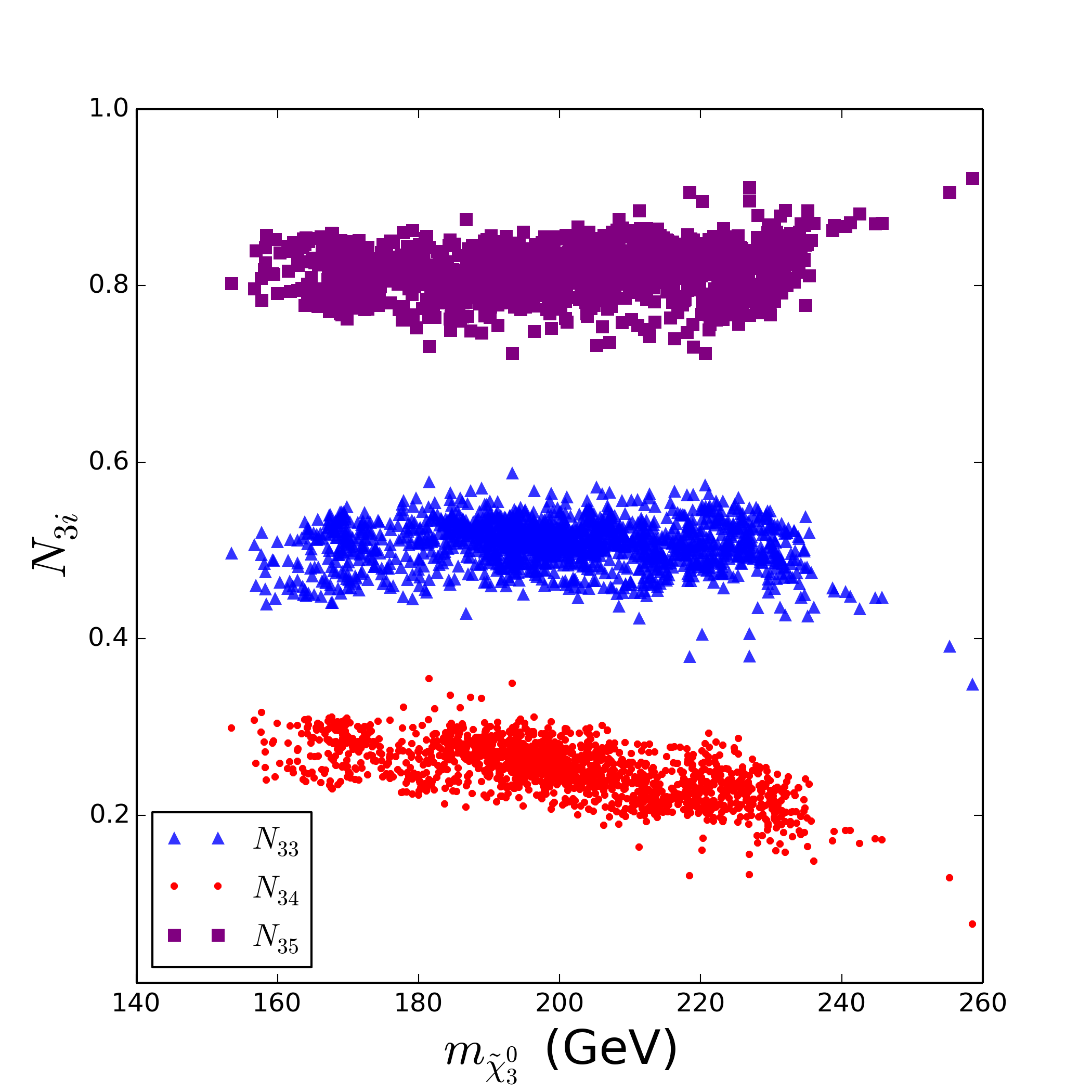}
\vspace{-0.4cm}
\caption{Components of the physical states  $\tilde{\chi}_1^0$, $\tilde{\chi}_2^0$ and $\tilde{\chi}_3^0$ in the NS scenario
with $\tilde{\chi}_1^0$ being Higgsino-dominated. Note that these results can be understood by Eq.(\ref{neutralino-mixing}).  \label{fig3}}
\end{figure}

In Fig.\ref{fig2}, we show the mass spectrum of $\tilde{\chi}_1^0$, $\tilde{\chi}_2^0$, $\tilde{\chi}_3^0$ and $\tilde{\chi}_1^\pm$ in our scenario.
This figure indicates that the mass splittings among the particles satisfy  $ 30 {\rm GeV} \lesssim \Delta_{\pm} \lesssim
70 {\rm GeV}$, $ 50 {\rm GeV} \lesssim  \Delta_{0} \lesssim  110 {\rm GeV}$ and $ 80 {\rm GeV} \lesssim
(m_{\tilde{\chi}_3^0} - m_{\tilde{\chi}_1^0}) \lesssim 160 {\rm GeV}$. We remind that these splittings are induced by
the strong mixings between Higgsinos and Singlino, and significantly larger than those among $\tilde{\chi}_1^0$, $\tilde{\chi}_2^0$ and
$\tilde{\chi}_1^\pm$ in the NS scenario of the MSSM \cite{Baer:2012uy} .  In Fig.\ref{fig3}, we show the field components of the states $\tilde{\chi}_1^0$,
$\tilde{\chi}_2^0$ and $\tilde{\chi}_3^0$ respectively. As is expected, the $\tilde{H}_u^0$, $\tilde{S}^0$ and $\tilde{H}_d^0$ components in
$\tilde{\chi}_1^0$ are comparable in magnitude with the largest one coming from the $\tilde{H}_u^0$ component. We emphasize again that the large Singlino component, i.e.
$N_{15} \sim 0.5$, can dilute the interactions of the Higgsino-dominated $\tilde{\chi}_1^0$ with other fields, and consequently $\tilde{\chi}_1^0$
can reach its right relic density. In this case, we checked that the main annihilation channels of $\tilde{\chi}_1^0$ in early universe
include $\tilde{\chi}_1^0 \tilde{\chi}_1^0 \to W^+ W^-, Z Z, Z h_1, h_1 h_1, h_1 h_2, q \bar{q}$.  As for $\tilde{\chi}_2^0$, its largest component
comes from either $\tilde{H}_d^0$ field (for most cases) or  $\tilde{H}_u^0$ field (in rare cases), and in general the two components are comparable, which can
be learned from the figure and also from Eq.(\ref{neutralino-mixing}). We checked that due to the spectrum and the mixings, the dominant decay of $\tilde{\chi}_1^\pm$ is
$\tilde{\chi}_1^\pm \to \tilde{\chi}_1^0 W^\ast$, and that of $\tilde{\chi}_2^0$ is usually $\tilde{\chi}_2^0 \to \tilde{\chi}_1^0 Z^{(\ast)}$.
By contrast, the possible dominant decay modes of $ \tilde{\chi}_3^0$ are rather rich, which include $\tilde{\chi}_1^0 Z^{(\ast)}, \tilde{\chi}_1^0 h_1, \tilde{\chi}_1^\pm W^{(\ast)} $.
Since $ \tilde{\chi}_3^0$ is Singlino-dominated, its production rate is rather low, and consequently its phenomenology is of less interest.

\begin{figure}[t]
\centering
\includegraphics[height=6cm,width=15cm]{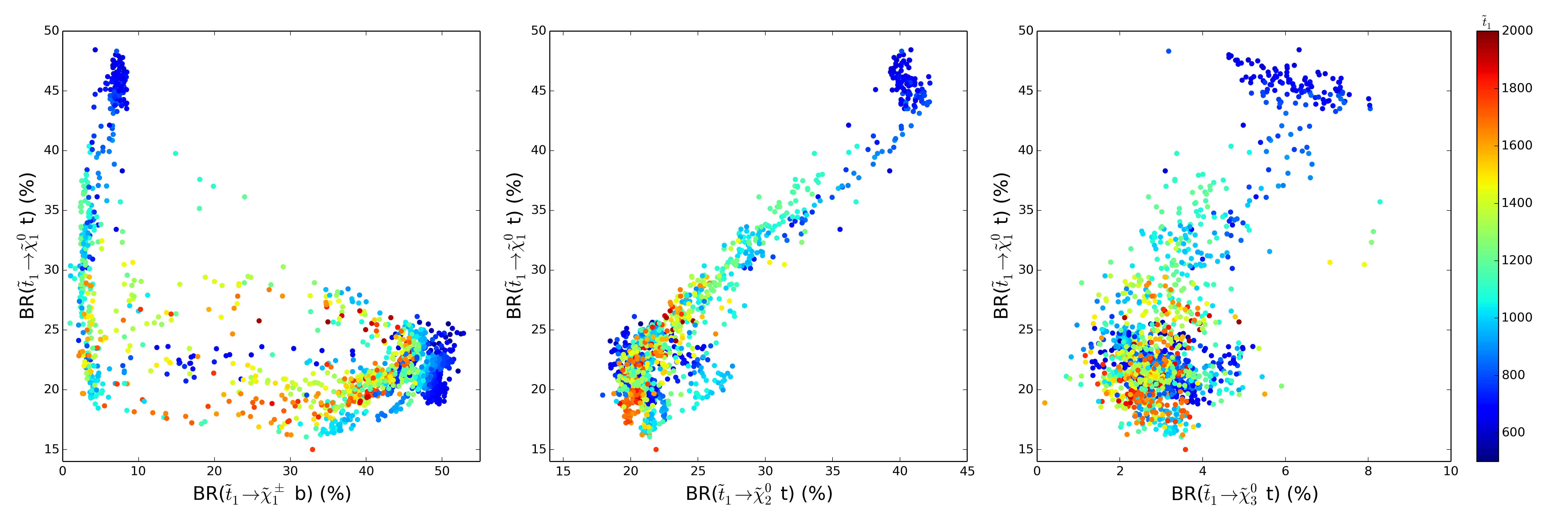} \\
\vspace{-0.4cm}
\caption{Correlations among the main decay modes of $\tilde{t}_1$ in the NS scenario
with $\tilde{\chi}_1^0$ being Higgsino-dominated.  Note that if $\tilde{t}_1$ is $\tilde{t}_R$ dominated
and meanwhile $|N_{14}| \simeq |N_{24}|$, we have $Br(\tilde{t}_1 \to \tilde{\chi}_1^+ b): Br(\tilde{t}_1 \to \tilde{\chi}_1^0 t):
Br(\tilde{t}_1 \to \tilde{\chi}_2^0 t) \simeq 2:1:1$. On the other hand, if $\tilde{t}_1$ is $\tilde{t}_L$ dominated,
$\tilde{t}_1$ prefers to decay into the Higgsino-dominated
$\tilde{\chi}_{1,2}^0$ with $Br(\tilde{t}_1 \to \tilde{\chi}_1^0 t) \simeq Br(\tilde{t}_1 \to \tilde{\chi}_2^0 t)$.  \label{fig4}}
\end{figure}

Next we turn to the properties of $\tilde{t}_1$. From the interactions of $\tilde{t}_1$ presented in \cite{MSSM-1}, one can infer that if $\tilde{t}_1$ is $\tilde{t}_R$ dominated
and meanwhile $|N_{14}| \simeq |N_{24}|$, the relation $Br(\tilde{t}_1 \to \tilde{\chi}_1^+ b): Br(\tilde{t}_1 \to \tilde{\chi}_1^0 t): Br(\tilde{t}_1 \to \tilde{\chi}_2^0 t) \simeq 2:1:1$
should hold. On the other hand, if $\tilde{t}_1$ is $\tilde{t}_L$ dominated, $\tilde{t}_1$ prefers to decay into the Higgsino-dominated
$\tilde{\chi}_{1,2}^0$ with $Br(\tilde{t}_1 \to \tilde{\chi}_1^0 t) \simeq Br(\tilde{t}_1 \to \tilde{\chi}_2^0 t)$.
These features are exhibited in Fig.\ref{fig4}, where we show the correlations between different decay rates of $\tilde{t}_1$ in our scenario.
From Fig.\ref{fig4}, one can also learn that the branching ratio of $\tilde{t}_1 \to \tilde{\chi}_3^0 t$
is less than $10\%$. This is because $\tilde{\chi}_3^0$ is Singlino-dominated and its $\tilde{H}_u^0$ component is small. Moreover, we note that in our scenario
$m_{\tilde{t}_1}$ is lower bounded by about $500 {\rm GeV}$, which is about $100 {\rm GeV}$ less than that in the NS scenario of the MSSM \cite{Kobakhidze:2015scd,Cao:2012rz,Han:2013kga}.
One reason for the difference is that  $\tilde{t}_1$ in our scenario has richer decay modes.

\section{Future detection of our scenario}

\subsection{Detection at 14 TeV LHC}
From the analysis in last section, one can learn that the NS scenario with
$\tilde{\chi}_1^0$ being Higgsino-dominated is characterized by predicting $\mu \leq 160 GeV$ and sizable
mass splittings among the Higgsino-dominated neutralinos and chargino, i.e. 30 GeV $\leq \Delta_{\pm} \leq $ 70 GeV and
50 GeV $\leq \Delta_{0} \leq $ 110 GeV. Although this kind of spectrum is allowed by the direct searches for the electroweakinos at LHC Run I,
it is expected to be tightly constrained at the upgraded LHC.

\begin{table}[th]
\small
\caption{Mass and decay information of the benchmark points P1, P2, P3 and P4 in our study.} \label{table-3}

\vspace{0.3cm}

\centering
\begin{tabular}{|c|c|c|c|c|c|c|c|}
\hline	
         & $m_{\tilde{\chi}_{1}^{0}}$  & $m_{\tilde{\chi}_{2}^{0}}$  & $m_{\tilde{\chi}_{3}^{0}}$ & $m_{\tilde{\chi}_{1}^{\pm}}$   & BR($\tilde{\chi}_{2}^{0}\to \tilde{\chi}_{1}^{0}Z^{(*)}$) & BR($\tilde{\chi}_{3}^{0}\to \tilde{\chi}_{1}^{0}Z^{(*)}$)& BR($\tilde{\chi}_{1}^{\pm}\to \tilde{\chi}_{1}^{0}W^*$)\\
\hline
P1    & 80.2   & 129.1 & 158.4 & 108.0 & 94.2\% & 9.08\% &100\%\\
\hline
P2    & 67.3   & 142.6 & 180.0 & 110.2 & 94.7\%&  7.55\%&100\%\\
\hline
P3    & 82.5   & 165.5 & 219.2 & 135.9 & 98.1\%& 26.3\%& 100\%\\
\hline
P4    & 74.9   & 193.4 & 220.6 & 147.6 & 96.2\%& 6.50\% & 100\%\\
\hline													
\end{tabular}
\end{table}

\begin{table}[t]
\caption{Cross sections of the four benchmark points in each bin of the signal region SR0$\tau$a, which are obtained from our simulations and
presented in the last four columns of this table. Quantities from the second column to the fifth column define the bins of
the SR0$\tau$a, and their physical meanings are explained in Appendix B.1.  The sixth column corresponds to the backgrounds of the bins, which
are taken from \cite{Cao:2015efs}. All quantities with mass dimension and cross sections are in units of GeV and fb respectively.}\label{table-4}

\vspace{0.3cm}

\centering
\begin{tabular}{|c|c|c|c|c|c|c|c|c|c|c|c|}
\hline
SR0$\tau$a & $m_{SFOS}$ & $m_T$  & $E_T^{miss}$ & $m_{3l}$ & Background  & P1    & P2   & P3    & P4   \\
\hline	
1	&	12-40	&	0-80	&	50-90	&	no	&	2.41	&	0.652 	&	0.423 	&	0.183 	&	0.007 		\\
\hline	
2	&	12-40	&	0-80	&	$>$90	&	no	&	0.45	&	0.273 	&	0.176 	&	0.108 	&	0.003 		\\
\hline	
3	&	12-40	&	$>$80	&	50-75	&	no	&	1	&	0.070 	&	0.054 	&	0.040 	&	0.001 		\\
\hline	
4	&	12-40	&	$>$80	&	$>$75	&	no	&	1.08	&	0.064 	&	0.074 	&	0.074 	&	0.008 		\\
\hline																				
5	&	40-60	&	0-80	&	50-75	&	yes	&	1.37	&	0.131 	&	0.365 	&	0.170 	&	0.006 		\\
\hline	
6	&	40-60	&	0-80	&	$>$75	&	no	&	0.76	&	0.119 	&	0.509 	&	0.302 	&	0.013 		\\
\hline	
7	&	40-60	&	$>$80	&	50-135	&	no	&	1.49	&	0.122 	&	0.240 	&	0.183 	&	0.011 		\\
\hline	
8	&	40-60	&	$>$80	&	$>$135	&	no	&	0.2	&	0.008 	&	0.022 	&	0.022 	&	0.002 		\\
\hline																				
9	&	60-81.2	&	0-80	&	50-75	&	yes	&	2.4	&	0.032 	&	0.156 	&	0.218 	&	0.040 		\\
\hline	
10	&	60-81.2	&	$>$80	&	50-75	&	no	&	1.51	&	0.027 	&	0.074 	&	0.087 	&	0.015 		\\
\hline	
11	&	60-81.2	&	0-110	&	$>$75	&	no	&	2.98	&	0.062 	&	0.312 	&	0.438 	&	0.094 		\\
\hline	
12	&	60-81.2	&	$>$110	&	$>$75	&	no	&	0.63	&	0.039 	&	0.072 	&	0.082 	&	0.017 		\\
\hline																				
13	&	81.2-101.2	&	0-110	&	50-90	&	yes	&	66.41	&	0.024 	&	0.415 	&	0.146 	&	0.870 		\\
\hline	
14	&	81.2-101.2	&	0-110	&	$>$90	&	no	&	21.62	&	0.016 	&	0.303 	&	0.107 	&	0.744 		\\
\hline	
15	&	81.2-101.2	&	$>$110	&	50-135	&	no	&	5.98	&	0.031 	&	0.086 	&	0.047 	&	0.157 		\\
\hline	
16	&	81.2-101.2	&	$>$110	&	$>$135	&	no	&	0.59	&	0.006 	&	0.018 	&	0.025 	&	0.032 		\\
\hline																				
17	&	$>$101.2	&	0-180	&	50-210	&	no	&	7.65	&	0.066 	&	0.136 	&	0.091 	&	0.032 		\\
\hline	
18	&	$>$101.2	&	$>$180	&	50-210	&	no	&	0.44	&	0.008 	&	0.026 	&	0.017 	&	0.001 		\\
\hline	
19	&	$>$101.2	&	0-120	&	$>$210	&	no	&	0.24	&	0.002 	&	0.003 	&	0.002 	&	0.005 		\\
\hline	
20	&	$>$101.2	&	$>$120	&	$>$210	&	no	&	0.09	&	0.002 	&	0.005 	&	0.007 	&	0.000 		\\
\hline																																		
\end{tabular}
\end{table}

\begin{table}[th]
\small
\caption{The best two signal bins and corresponding significances for the benchmark points with 30 fb$^{-1}$ and 300 fb$^{-1}$ integrated luminosity data respectively
at 14 TeV LHC.}\label{table-5}

\vspace{0.3cm}

\centering
\begin{tabular}{|c|l|l|l|l|l|l|l|l|}
\hline	
         &\multicolumn{2}{c}{30 fb$^{-1}$} 					& \multicolumn{2}{|c|}{300 fb$^{-1}$}\\
\hline
P1    & $S$(bin2) =  2.09 & $S$(bin1) = 1.75 		&$S$(bin2) = 4.59	&$S$(bin1) = 2.54  \\
\hline
P2    & $S$(bin6) =  2.88 & $S$(bin5) =1.44 		&$S$(bin6) = 5.58 	&$S$(bin2) = 2.96  \\
\hline
P3    & $S$(bin6) =   1.71 & $S$(bin11) = 1.01 		&$S$(bin6) = 3.31 	&$S$(bin2) = 1. 82  \\
\hline
P4    & $S$(bin14) =  0.32 & $S$(bin15) = 0.21 	 	&$S$(bin16) = 0.42 	&$S$(bin14) = 0.34  \\
\hline													
\end{tabular}
\end{table}

We investigate this issue by considering the neutralino and chargino associated production processes at 14 TeV LHC. For simplicity we adopt 4 benchmark points listed in
Table \ref{table-3}, which are discriminated by the values of $\mu$ (or equivalently $m_{\tilde{\chi}_1^\pm}$), $\Delta_{0}$ and $\Delta_{\pm}$.
Since $\tilde{\chi}_1^\pm$ for these points decays into $\tilde{\chi}_1^0$ plus an off-shell $W$ boson, and $\tilde{\chi}_2^0$ decays mainly
into $\tilde{\chi}_1^0$ plus a $Z$ boson (on-shell or off-shell),
the signal region SR0${\tau a}$ in the ATLAS direct searches for electroweakinos by trileptons and large $E_T^{miss}$ signal \cite{Aad:2014nua}, 
which was proposed in the analysis \cite{Baer:1985at}
and also briefly introduced in the appendix of this work, is most pertinent to explore those points. In our analysis, we simulate
the processes $ p p \to \tilde{\chi}_1^\pm \tilde{\chi}_2^0, \tilde{\chi}_1^\pm \tilde{\chi}_3^0 \to 3 l + E_T^{miss} $ to get their summed rate
in each bin of the signal region, and present the result in the last four columns  of Table \ref{table-4}.
We also present in the table the  backgrounds of the bins at 14 TeV LHC, which were obtained by detailed simulations done in \cite{Cao:2015efs}.
With these results, we evaluate the significance $S = s /\sqrt{b+(\epsilon b)^2 }$ for each bin, where $s$ and $b$ correspond to the number of signal
and background events and $\epsilon=10\%$ is the assumed systematical uncertainty of the backgrounds.
Assuming 30 fb$^{-1}$ and 300 fb$^{-1}$ integrated luminosity data  at 14 TeV LHC, we present the best two signal  bins
and corresponding expected significances for P1, P2, P3 and P4 in Table \ref{table-5}. This table reveals following information:
\begin{itemize}
\item With 30 fb$^{-1}$ integrated luminosity data, P1 and P2 can be excluded at 2$\sigma$ confidence level, and with 300 fb$^{-1}$ data P3 can also be excluded.
  In any case, the point P4 is hard to be excluded.
\item For each point, which signal bin is best for exclusion depends on the mass splittings among the neutralinos and chargino.
 For example, since $\Delta_{\pm}< m_W$ for all the four points, the most effective bins usually require $m_T<$ 80 or 110 GeV. For points P1, P2 and P3,
 the bins satisfying $m_{SFOS} < m_Z$ are preferred for exclusion since $\Delta_0 < m_Z$, and by contrast the bins with $|m_{SFOS}-m_Z|<$ 10 GeV
 (such as bins 14 and 15) are favored by point P4 since in this case $\tilde{\chi}_2^0$ can decay into an on-shell $Z$ boson. Note that for bins 14 and 15, the backgrounds are
 relatively large, and that is why the point P4 can not be excluded at 14 TeV LHC after including the systematic uncertainties.
\item With 300 fb$^{-1}$ data, the point P2 can be discovered at 14 TeV LHC. This is partially because $\tilde{\chi}_2^0$ and $\tilde{\chi}_1^\pm$ are relatively light
so that the rate of their associated production is large, and also because they have sizable mass splittings from $\tilde{\chi}_1^0$ to result in moderately
energetic decay products.
\item Since the bins in the SR0$\tau$a are disjoint, in principle their results can be statistically combined to maximize the significance. We did this, but we found that
the improvement is not significant.
\end{itemize}

Since the four benchmark points stand for the typical situation in our scenario,
we conclude that the future LHC experiments can exclude most part of the parameter space for the scenario,
and consequently the fine-tuning of the NMSSM can be pushed to higher level. We will discuss such an issue  extensively
in our forthcoming work.

\begin{figure}
\centering
\includegraphics[width=17cm]{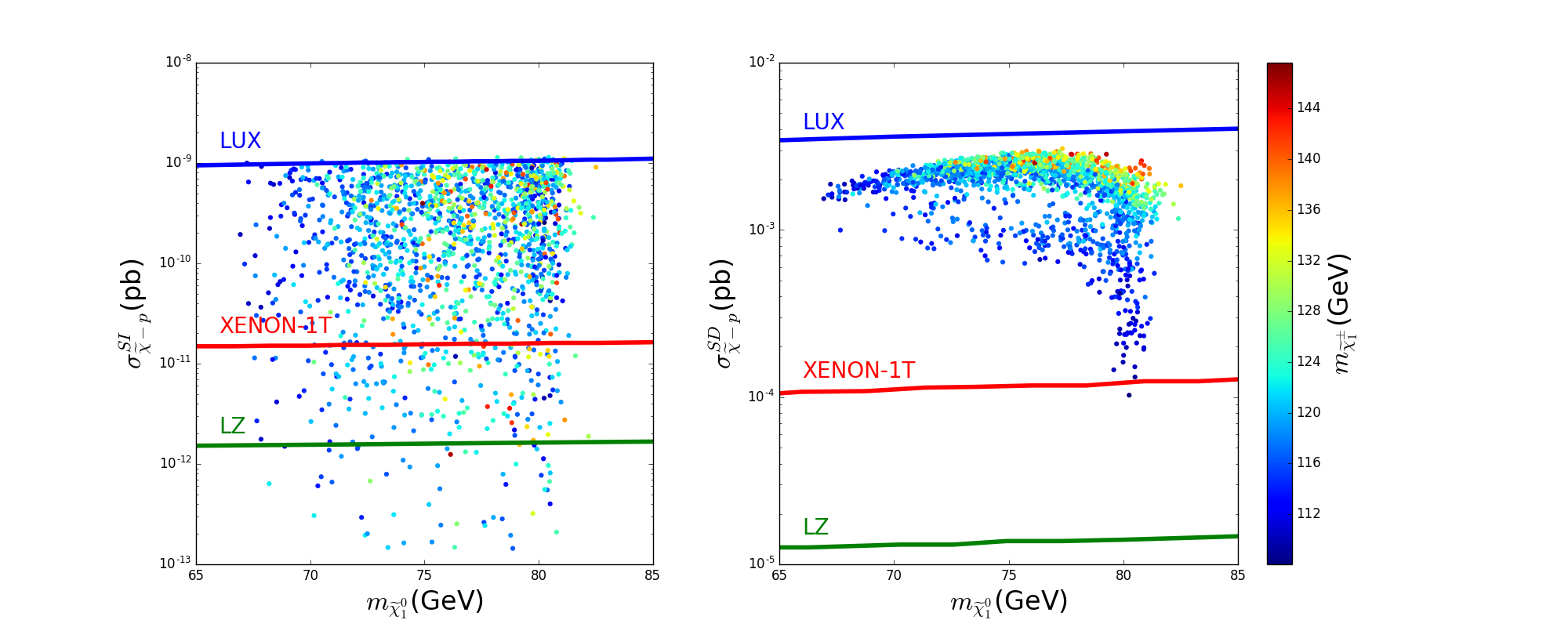}

\vspace{-0.5cm}

\caption{Spin-independent (SI) and Spin-dependent (SD) DM-nucleon scattering cross sections versus DM mass
in the NS scenario with $\tilde{\chi}_1^0$ being Higgsino-dominated. Capabilities of future DM direct detection experiments
in detecting the scattering are also plotted. \label{fig5} }
\end{figure}

\subsection{Dark Matter Direct Search}

Since $\tilde{\chi}_1^0$ as the DM candidate has quite large $\tilde{H}_u^0$, $\tilde{H}_d^0$ and $\tilde{S}^0$ components in our scenario, its interactions with the Higgs bosons and $Z$ boson
are unsuppressed. As a result, the spin-independent (SI) and spin-dependent (SD) cross sections of the $\tilde{\chi}_1^0$-nucleon scattering are sizable, and may reach
the sensitivities of future DM direct detection experiments such as XENON-1T and LZ-7.2T experiments \cite{direct:future}. In this section, we investigate such an issue.
In Fig.\ref{fig5}, we project the samples of our scenario on $m_{\tilde{\chi}_1^0}-\sigma^{SI}_{\tilde{\chi}-p}$ and $m_{\tilde{\chi}_1^0}-\sigma^{SD}_{\tilde{\chi}-p}$ planes with
$\sigma^{SI}_{\tilde{\chi}-p}$ and $\sigma^{SD}_{\tilde{\chi}-p}$ denoting the SI and SD cross sections respectively.
The blue lines, red lines and green lines are the sensitivities of LUX, XENON-1T and LZ experiments respectively.

For the SI cross section, one can learn from Fig.\ref{fig5} that the future XENON-1T experiment is able to probe a large portion of the samples, and the
LZ experiment can test even more. Anyhow, there still exist some samples remaining untouched by these future experiments.
We numerically checked that for the untouched samples, there exists rather strong cancelation among the contributions induced by different CP-even Higgs bosons.
On the other hand, the story for $\sigma^{SD}_{\tilde{\chi}-p}$  is quite different. From the right panel of Fig.\ref{fig5} one can see that the future XENON-1T experiment
can test nearly all of the samples, let alone the more sensitive LZ experiment. The underlying reason is that in the NMSSM with heavy sfermions,
the SD cross section gets contribution mainly from the $t$-channel $Z$-mediated diagram.  As a result, the size of the cross section is determined by the
$Z \tilde{\chi}_1^0 \tilde{\chi}_1^0$ coupling, which is given by
\begin{eqnarray}
g_{Z \tilde{\chi}_1^0 \tilde{\chi}_1^0} = \frac{m_Z}{\sqrt{2} v} ( N_{13}^2 - N_{14}^2 ) \sim 0.2 \times \frac{m_Z}{\sqrt{2} v}
\end{eqnarray}
In the last step of the equation, we have used the information of $N_{13}$ and $N_{14}$ presented in Fig.\ref{fig3}. Moreover,
it is interesting to see that although the benchmark point P4 in Table \ref{table-3} is hard to be excluded by sparticle direct search
experiments at the LHC, its SD cross section is quite large and so the point will be tested by future dark matter direct search experiments.

In getting Fig.\ref{fig5}, we use the package micrOMEGAs \cite{micrOMEGA} to calculate the cross sections. We choose
its default setting $\sigma_{\pi N} = 34  {\rm MeV}$ and $\sigma_0 = 42  {\rm MeV}$ as input. We checked that
if we take $\sigma_{\pi N} = 59  {\rm MeV}$ from \cite{SI-piN} and $\sigma_0 = 58  {\rm MeV}$ from \cite{SI-pi0-1,SI-pi0-2,SI-pi0-3},
the SI cross section will be enhanced by a factor from $20\%$ to $40\%$, and this does not
affect the conclusions presented in this work.

\section{Conclusions}

With the great discovery of the 125 GeV Higgs boson and the increased bounds on sparticle masses at LHC Run I, the NS scenario in MSSM
has become theoretically unsatisfactory. By contrast, the situation may be improved greatly in the NMSSM. This motivates us to scrutinize the
impact of the direct searches for SUSY at LHC Run I on the naturalness of the NMSSM.

We start our study by scanning the vast parameter space of the NMSSM to get the region where the fine tuning measures
$\Delta_Z$ and $\Delta_h$ at electroweak scale are less than about 50. In this process, we considered various experimental
constraints such as DM relic density, LUX limits on the scattering rate of DM with nucleon and the 125 GeV Higgs data
on the model. We classify the surviving samples into four types: for Type I samples, $h_1$ corresponds to the SM-like Higgs
boson and $\tilde{\chi}_1^0$ is Bino-dominated, while for Type II, III and IV samples, $h_2$ acts as the SM-like Higgs boson
with $\tilde{\chi}_1^0$ being Bino-, Singlino- and Higgsino-dominated respectively.
After these preparations, we specially study the influence of the direct searches for SUSY on the samples.
We implement the direct search constraints by the packages FastLim and SModelS and also by simulating
the electroweakino and stop production processes. Our results indicate that although the direct search experiments
are effective in excluding the samples, the parameter intervals for the region and also the minimum reaches of
$\Delta_Z$ and $\Delta_h$ are scarcely changed by the constraints, which implies that,
contrary to general belief, the fine tuning of the NMSSM does not get worse after LHC Run I.
Our results also indicate that the lowest fine-tuning comes from type III and type IV samples, for which $\Delta_Z$ and $\Delta_h$ may be as low as
about 2 without conflicting with any experimental constraints.

Considering that the type IV samples were scarcely studied in previous literatures and that they have similar phenomenology to that of
the NS scenario in the MSSM, we investigate the essential features of this kind of samples. We find
that they are characterized by strong mixings between Higgsinos and Singlino in forming mass eigenstates called neutralinos.
As a result, the lightest neutralino $\tilde{\chi}_1^0$ as the DM candidate has significant Singlino component so that it
can easily reach the measured DM relic density, and meanwhile the mass splittings among the Higgsino-dominated particles $\tilde{\chi}_1^0$,
$\tilde{\chi}_2^0$ and $\tilde{\chi}_1^\pm$ are usually larger than $30 {\rm GeV}$. These features make
the samples rather special. For example, we show that due to the rich decay products of the lighter scalar top quark, its lower mass bound
is decreased by about 100GeV in comparison with that in the NS scenario of the MSSM,
and that the neutralino-chargino sector of the samples can be readily tested either through searching for $3 l + E_T^{miss}$ signal at
14 TeV LHC or through future dark matter direct detection experiments.

In summary, we conclude that so far the fine tuning of the NMSSM is scarcely affected by the direct searches for SUSY at LHC Run I, and it
can still predict $Z$ boson mass and the SM-like Higgs boson mass in a natural way.  This conclusion, however, may be altered
by the upcoming 14 TeV LHC experiments and DM matter direct search experiments.

\section*{Acknowledgement}

This work was supported by the National Natural Science Foundation
of China (NNSFC) under grant No. 11575053, 11275245 and 11305050.

\appendix

\section{Fastlim and SModelS}

\begin{table}[htbp]
\centering
\caption{Experiments in the Fastlim database.}
\label{t-fastlim}
\begin{footnotesize}
\begin{tabular}{|c|c|c|c|c|}
\hline
Name & Description & $\sqrt{s}$ & $\mathcal{L}_{\text{int}}(\text{fb}^{-1})$ & Ref. \\ \hline
ATLAS-CONF-2013-062 & 1-2 leptons + 3-6 jets + $E_T^{miss}$ (squarks and gluino) & \multirow{8}{*}{8TeV} & 20.3 &  \cite{atlas:2013062} \\ \cline{1-2} \cline{4-5}
ATLAS-CONF-2013-061 & 0-1 lepton + $\ge$3 b-jets + $E_T^{miss}$ (3rd gen. squarks) &  & 20.1 & \cite{atlas:2013061} \\ \cline{1-2} \cline{4-5}
ATLAS-CONF-2013-054 & 0 lepton + $\ge$ 7-10 jets + $E_T^{miss}$ (squarks and gluino) &  & 20.3 &  \cite{atlas:2013054} \\ \cline{1-2} \cline{4-5}
ATLAS-CONF-2013-053 & 0 lepton + 2 b-jets + $E_T^{miss}$ (sbottom and stop) &  & 20.1 &  \cite{atlas:2013053} \\ \cline{1-2} \cline{4-5}
ATLAS-CONF-2013-048 & 2 leptons (+ jets) + $E_T^{miss}$ (stop) &  & 20.3 &  \cite{atlas:2013048} \\ \cline{1-2} \cline{4-5}
ATLAS-CONF-2013-047 & 0 lepton + 2-6 jets + $E_T^{miss}$ (squarks and gluino) &  & 20.3 &  \cite{atlas:2013047} \\ \cline{1-2} \cline{4-5}
ATLAS-CONF-2013-037 & 1 lepton + 4(1 b-)jets + $E_T^{miss}$ (stop) &  & 20.7 &  \cite{atlas:2013037} \\ \cline{1-2} \cline{4-5}
ATLAS-CONF-2013-024 & 0 lepton + (2 b-)jets + $E_T^{miss}$ (stop) &  & 20.5 &  \cite{atlas:2013024} \\ \hline
\end{tabular}
\end{footnotesize}
\end{table}

In this section, we  briefly introduce the packages Fastlim \cite{Papucci:2014rja}  and SModelS \cite{Kraml:2013mwa}, which can be used to implement the constraints from the direct
searches for sparticles at LHC Run I in an easy and fast way.

\subsection{Fastlim}

Fastlim is a package that limits SUSY parameter space using the LHC-8TeV data. It incorporates in its database cut efficiency tables for different sparticle
production processes in simplified model framework. Now it supports the processes of gluino pair production and
third generation squark pair productions, and it involves many decay modes of gluino and the squarks, such as $\tilde{g}\to \tilde{t}_{1,2}^\ast t \to \tilde{\chi}_1^0  t \bar{t}$
or $\tilde{g}\to q \bar{q}  \tilde{\chi}_1^0$ and $\tilde{b}_{1,2} \to b \tilde{\chi}_1^0$ where $q$ stands for the first two generation quarks.
Any processes that contain the same final states at the detector level are combined by Fastlim to improve the signal significance.

In this work, we only use the experiments of searching for gluino and squarks, which are listed in Table \ref{t-fastlim}. This is because that
Fastlim gives better limits for strong SUSY productions in comparison with the package SModelS \cite{Kraml:2013mwa}.

\subsection{SModelS}

SModelS  have the same function as that of Fastlim, but it could give better limits for slepton productions and electroweakino productions.
SModelS has implemented the information about following experiments:
\begin{itemize}
\item Direct slepton searches (ATLAS): ATLAS-CONF-2013-049 \cite{atlas:2013049}.
\item Direct slepton searches (CMS): SUS-12-022 \cite{cms:12022}, SUS-13-006 \cite{cms:13006}.
\item Electroweakino searches (ATLAS): ATLAS-CONF-2013-028 \cite{atlas:2013028}, ATLAS-CONF-2013-035 \cite{atlas:2013035},
            ATLAS-CONF-2013-036 \cite{atlas:2013036}, ATLAS-CONF-2013-093 \cite{atlas:2013093}.
\item Electroweakino searches (CMS): SUS-12-022 \cite{cms:12022}, SUS-13-006 \cite{cms:13006}, SUS-13-017 \cite{cms:13017},
\end{itemize}
and it contains the cross section limits at 95\% confidence level for above analyses.  If one parameter point predicts
cross sections larger than those presented in the analyses, it will be excluded. Otherwise, it is allowed.

Note that the efficiencies or the upper bounds on SUSY signals in the database of the two packages are usually based on certain assumptions, which may not be applied
to some parameter points encountered in our scan. In this case, the encountered point is considered to be experimentally allowed. Also note that
the packages are based on the preliminary analyses of the ATLAS and CMS groups, which were done in 2013. Given that most of these analyses have been updated in past two years,
a more powerful exclusion capability may be obtained if one repeats the updated analyses by detailed Monte Carlo simulation.  Anyhow, these two packages
can serve as useful tools to exclude some SUSY parameter points.

\section{Details of our simulation}

In our study, any point that passed Fastlim and SModelS is further tested by simulations to see whether
it survives the constraints from the direct search experiments. In detail, we first use MadGraph/MadEvent \cite{Alwall:2011uj}
to generate parton level events for certain sparticle production processes, and feed them into Pythia \cite{Sjostrand:2006za}
for parton showering and hadronization. Then we use the package CheckMATE \cite{Drees:2013wra} where
a well-tuned Delphes \cite{deFavereau:2013fsa} is provided for the detector simulation and analyses.
We define $R\equiv \text{max}\left\{S_i/S_{i, obs}^{95}\right\}$ to decide whether
the point survives the analysis, where $S_i$ stands for the simulated signal events in the $i$th signal region of the analysis,
and $S_{i, obs}^{95}$ represents the $95\%$ C.L. upper limit of the event number in the signal region.
If $R>1$, the parameter point is excluded by the analysis and otherwise it is allowed.

Since the fine tuning of the NMSSM at the electroweak scale is mainly affected by its chargino-neutralino sector and stop sector,
we repeat by simulation the ATLAS analyses in \cite{Aad:2014vma}, \cite{Aad:2014nua}, \cite{Aad:2014kra} and \cite{Aad:2014mha}.
All these analyses are based on 20.3fb$^{-1}$ data at the LHC-8TeV with the former two presenting so far the strictest limits on electroweakino
productions, and the latter two providing the tightest constraints on the pair productions of third generation squarks.
In the following, we briefly introduce these analyses.

\subsection{Search for Electroweakino at the LHC}

The analysis \cite{Aad:2014vma}  targets final states with two leptons and large $E_T^{miss}$. In our simulation of the analysis, we
mainly focus on the signal region named "SR-Zjets", which is specifically designed for the process
$p p \to \tilde{\chi}_{2}^0\tilde{\chi}_1^{\pm}\to Z\tilde{\chi}_1^0 W\tilde{\chi}_1^0
\to \ell\ell\tilde{\chi}_1^0 q q \tilde{\chi}_1^0$. This signal region requires that the two leading
leptons should be same flavor but opposite sign (SFOS), and that their invariant mass locates at $Z$-peak.
As was pointed out in \cite{Aad:2014vma}, it provides the strongest constraints on the chargino-neutralino
sector among the anlayses with dilepton final state.

The analysis in \cite{Aad:2014nua} also searches for electroweakino productions
but with final states of three leptons and large $E_T^{miss}$.
Here we concentrate on the signal region named "SR0$\tau$a" which is optimized for processes
$p p \to \tilde{\chi}_{2}^0\tilde{\chi}_1^{\pm}\to Z^{(\ast)} \tilde{\chi}_1^0 W^{(\ast)} \tilde{\chi}_1^0
 \to \ell\ell\tilde{\chi}_1^0 \ell\nu \tilde{\chi}_1^0$. This region requires a pair of SFOS leptons in its signal,
and utilizes the transverse mass $m_T=\sqrt{2\left|\vec p_T^{~\ell}\right|\left|\vec E_T^{miss}\right|
 -2\vec p_T^{~\ell} \cdot \vec E_T^{miss}}$ (here $p_T^l$ is the transverse momentum of the lepton not forming
 the SFOS lepton pair) to suppress the SM background.
It considers twenty bins, which are categorized by the SFOS leptons' invariant mass, $m_T$ and $E_T^{miss}$, to maximize its sensitivity
to different mass spectrums of $\tilde{\chi}_{2}$, $\tilde{\chi}_1^{\pm}$
and $\tilde{\chi}_1^0$.

We remind that in the NS scenarios of the NMSSM, all $\tilde{\chi}_{i}^0\tilde{\chi}_1^{\pm}$ associated production processes with $i=2,3,4,5$
may contribute sizably to the trilepton signal, so in our simulation we include all these contributions. By contrast,
SModelS does not combine processes that have the same final states at the detector level. Moreover, in our analysis
we combine the signal region "SR-Zjets"  with the bins in "SR0$\tau$a" to maximize the discovery
significance by the $CL_s$ method in RooStats as we did in \cite{Cao:2015efs}.

\subsection{Search for stops at the LHC}

In the NS scenario of the NMSSM, $\tilde{t}_{1}$ usually decays like $\tilde{t}_1 \to \tilde{\chi}_{2,3}^0 t \to \tilde{\chi}_{1}^0 Z^{(\ast)} t$
or $\tilde{t}_1 \to \tilde{\chi}_{1}^+ b \to \tilde{\chi}_{1}^0 W^{(\ast)} b$. Considering that these two-step topologies haven't
been included in the database of Fastlim, we repeat recent ATLAS analyses on stop pair productions, which were presented
in \cite{Aad:2014kra} and \cite{Aad:2014mha}.

The analysis in \cite{Aad:2014kra} searches for stops in final states containing exactly one isolated lepton,
at least two jets and a large $E_T^{miss}$. It contains fifteen signal regions targeting a large number of stop
pair production scenarios, where stop may decay like  $\tilde{t}_1\to t\tilde{\chi}_1^0$,
$\tilde{t}_1\to bW^{(\ast)}\tilde{\chi}_1^0$ and $\tilde{t}_1\to b\tilde{\chi}_1^{\pm}\to b W^{(\ast)}\tilde{\chi}_1^0$.
Especially, nine of these signal regions are designed for the latter decay with
different mass relations among $\tilde{t}_1$, $\tilde{\chi}_1^{\pm}$ and $\tilde{\chi}_1^0$, which determines the kinematic
properties of the process. These nine regions are particularly relevant to our study.
In general, the signal regions are discriminated by different kinematic cuts on the leptons and (b-)jets, $E_T^{miss}$, $m_T$, b-jet
multiplicity and the asymmetric transverse mass $am_{T2}$ \cite{Bai:2012gs}.

The analysis in \cite{Aad:2014mha} targets the process $p p \to \tilde{t}_2^\ast \tilde{t}_2$ with
$\tilde{t}_2\to\tilde{t}_1 Z\to\tilde{\chi}_1^0 Z t$, and it searches for the signal that contains a SFOS
pair of leptons with their invariant mass near $m_Z$, at least one b-jet and a large $E_T^{miss}$. Obviously,
if $\tilde{t}_2$ is significantly heavier than $\tilde{t}_1$, the $Z$ boson is highly boosted. As a result,
the transverse momentum of the dilepton system $p_T(\ell\ell)$ tends to be high and the azimuthal
separation $\Delta \phi({\ell\ell})$ prefers to be low. On the other side, if the mass splitting between
$\tilde{t}_1$ and $\tilde{\chi}_1^0$ is large, high jet multiplicity is expected. These facts motivates
physicists to define five signal regions, which are called "SR2(A,B,C)" and  "SR3(A,B)" respectively, by the
number of leptons in the signal, the jet multiplicity and whether the Z boson is boosted. For example,
the signal regions "SR2(A,B,C)" require that the signal events contain exactly two signal leptons
and the $Z$ boson is boosted. The regions "SR2A" and "SR2B" are optimized for low jet multiplicity,
while the region "SR2C" is designed for high jet multiplicity case.


\end{document}